\documentclass[sigconf,balance=false]{acmart}
\usepackage{popets}

\acmYear{YYYY}
\acmVolume{YYYY}
\acmNumber{X}
\acmDOI{XXXXXXX.XXXXXXX}
\acmISBN{}
\acmConference{}
\usepackage{xspace}
\newcommand{\benchmark}{{\sc GenAIPABench}\xspace}
\usepackage{algorithm}
\usepackage{float}  
\newfloat{algorithm}{t}{lop}
\usepackage[absolute,overlay]{textpos}
\usepackage{lipsum}
\usepackage[noend]{algpseudocode}
\usepackage[utf8]{inputenc} 
\usepackage[T1]{fontenc}    
\usepackage{hyperref}       
\usepackage{url}            

\usepackage{amsfonts}       
\usepackage{nicefrac}       
\usepackage{microtype}      
\usepackage{lipsum}         
\usepackage{graphicx}
\usepackage{doi}
\usepackage{amsmath}
\usepackage{cleveref} 
\usepackage[skip=2pt]{caption} 
\usepackage{subcaption}

\usepackage{multirow, makecell} 
\usepackage{booktabs}
\usepackage{color,soul}
\usepackage{todonotes}

\usepackage{subcaption}
\usepackage{xspace}

\usepackage[title]{appendix}
 
\begin{document}

\title{\benchmark: A Benchmark for \\ Generative AI-based Privacy Assistants}


\author{
    Aamir Hamid\textsuperscript{1,*},
    Hemanth Reddy Samidi\textsuperscript{1},
    Tim Finin\textsuperscript{1}, 
    Primal Pappachan\textsuperscript{2,†},
    Roberto Yus\textsuperscript{1}
    \\
    \textsuperscript{1}University of Maryland, Baltimore County, \textsuperscript{2}Portland State University
    \\
    \{ahamid2, finin, hsamidi1, ryus\}@umbc.edu, primal@pdx.edu
}




\begin{abstract}
Privacy policies of websites are often lengthy and intricate. \textit{Privacy assistants} assist in simplifying policies and making them more accessible and user-friendly. The emergence of generative AI (genAI) offers new opportunities to build privacy assistants that can answer users' questions about privacy policies. However, genAI's reliability is a concern due to its potential for producing inaccurate information. This study introduces \benchmark, a benchmark for evaluating Generative AI-based Privacy Assistants (GenAIPAs). \benchmark includes: 1) A set of questions about privacy policies and data protection regulations, with annotated answers for various organizations and regulations; 2) Metrics to assess the accuracy, relevance, and consistency of responses; and 3) A tool for generating prompts to introduce privacy documents and varied privacy questions to test system robustness. We evaluated three leading genAI systems—ChatGPT-4, Bard, and Bing AI—using \benchmark to gauge their effectiveness as GenAIPAs. Our results demonstrate significant promise in genAI capabilities in the privacy domain while also highlighting challenges in managing complex queries, ensuring consistency, and verifying source accuracy.
\end{abstract}

\maketitle

\section{Introduction}

In today's digital landscape, effectively managing and protecting personal information is crucial for both individuals and organizations. Data privacy has become a central issue, highlighting the need for strong privacy regulations. These regulations, including the EU's GDPR and California's CCPA, are enforcing strict guidelines to ensure the protection of user data against misuse or unauthorized access. One of the common requests to be compliant with the regulations is for organizations to provide users with information about how their data is managed in the form of privacy policies. However, both privacy policies and regulations often suffer from complexity~\cite{voigt2017eu,greenberg2020california, solove2013nothing,obar2018biggest}, making it difficult for users to comprehend their rights and the protections in place for their privacy.


To address privacy concerns, the concept of a \textit{privacy assistant} has been developed. These assistants, utilizing insights from privacy policy analysis, transform complex policies into accessible, user-friendly information and aid users in managing their data privacy more effectively~\cite{langheinrich2001privacy,ackerman2001privacy}. Privacy assistants come in various forms such as software applications, chatbots, and browser extensions. AI, with its capability to process vast data, adapt to user needs, and offer tailored recommendations~\cite{cavoukian2010privacy}, is particularly effective in privacy management. Research in this field includes developing AI tools for summarizing privacy policies~\cite{wilson2016summarizing}, providing personalized privacy recommendations~\cite{knijnenburg2013personalized}, and conducting privacy risk analyses~\cite{zhang2016privacy}.


The emergence of Large Language Models (LLMs) such as GPT~\cite{radford2018improving}, Llama~\cite{belinkov2020lama}, and BERT~\cite{devlin2018bert} represents a significant advancement in generative AI. These models excel in generating human-like text, having been trained on vast datasets. GPT-4.0, the latest version at the time of this writing, is a leader among LLMs, trained on trillions of tokens from the Internet, demonstrating exceptional contextual understanding and response accuracy~\cite{radford2021gpt}. Advanced chatbots like ChatGPT have also been developed using these models~\cite{radford2019language}. These genAI models and chatbots are increasingly being applied in domain-specific tasks, paving the way for a new generation of AI personal assistants. LLM-based chatbots, for instance, have shown great promise in various fields including customer support~\cite{gao2019neural}, healthcare~\cite{bickmore2005establishing}, personal finance management~\cite{li2017alice}, mental health support~\cite{fitzpatrick2017delivering}, and education~\cite{winkler2018evaluating}. Considering the critical importance of privacy and the challenges users face in understanding privacy policies, this trend suggests the potential emergence of highly efficient and reliable generative AI privacy assistants (to which we will refer in the following as \textit{GenAIPAs}).

While genAI features are promising, several challenges persist. The accuracy of LLM-generated responses is often questioned due to their propensity to produce ``glitches'' or incorrect information, impacting their trustworthiness~\cite{wang2021truth, schick2021s, bender2021dangers}. They may also generate misleading or erroneous references, further compromising their credibility. A recent study~\cite{chen2023chatgpts} highlights the need for a robust benchmark system for LLMs like GPT-3.5 and GPT-4 to ensure consistent performance evaluation and quality control and to promote transparency and accountability. The complexity of evaluating LLMs and genAI arises from their training on extensive datasets and their capability to produce text akin to human writing. A range of evaluation metrics such as F1, BLEU, ROUGE, METEOR scores, Adversarial evaluation, and CIDEr~\cite{ge2023openagi,kang2023llms,liu2023agentbench,bang2023multitask,zhang2023benchmarking} have been suggested, yet there is no single universally accepted metric due to domain-specific evaluation needs. In particular, privacy evaluation faces unique challenges, including the lack of clear ground truth, multidimensional objectives like data minimization and user consent, and the subjective nature of user perception, which often diverges from technical metrics. Hence,  while genAI systems have been evaluated in sectors like healthcare, finance, and even mental health, up to the authors' knowledge, they have not been evaluated in the privacy domain. This lack of focus on privacy-related aspects could leave users vulnerable to various risks, such as making misinformed decisions when sharing data with an online service, underscoring the urgent need for comprehensive evaluations in this field.

We have designed the \benchmark benchmark to evaluate genAI-enabled privacy assistants, focusing on diverse tasks in areas such as transparency, user control, data minimization, security, and encryption. The benchmark includes: 1) A selected corpus of privacy policies and regulations; 2) Policy-related questions sourced from FAQs, online forums, and direct user inquiries, accompanied by annotated answers; 3) Metrics to assess GenAIPA responses for relevance, accuracy, clarity, completeness, and reference; and 4) An evaluator tool that applies these metrics to gauge GenAIPA performance\footnote{Note that the benchmark's content is only in English.}. The main contributions of this paper are:
\begin{itemize}
\item The introduction of the first benchmark, to our knowledge, for evaluating GenAIPAs.
\item The assessment of three popular genAI chatbots (ChatGPT, BARD, and Bing Chat) using \benchmark.
\item An analysis of the results, highlighting challenges and opportunities in the development of GenAIPAs.
\end{itemize}

The rest of the paper is structured as follows. In Section~\ref{sect:related_work}, we review the state of the art on privacy benchmarking and genAI evaluation. In Section~\ref{sect:benchmark}, we introduce the benchmark. In Section~\ref{sect:questions} and Section~\ref{sect:metrics}, we detail \benchmark's question corpus and metrics, respectively. In Section~\ref{sect:results}, we present the experiments performed using \benchmark. In Section~\ref{sect:discussion}, we provide a discussion on challenges and opportunities. Finally, Section~\ref{sect:future} concludes the paper and presents directions for future research.

\section{Related Work}
\label{sect:related_work}
Since our benchmark is the first developed to assess the performance of GenAIPAs, we survey previous work on privacy benchmarks and on benchmarking general-purpose genAI systems as well as general-purpose question-answering systems.

\textbf{Privacy Benchmarks.} The growing interest in privacy benchmarks and evaluation frameworks has led to innovative projects to enhance the effectiveness, usability, and transparency of privacy policies and language models. For example, the PrivacyQA Project~\cite{Ravichander2019QuestionAF} developed a corpus containing 1,750 QAs on mobile app privacy policies, enriched with over 3,500 expert annotations, to improve user awareness and selective exploration of privacy issues. Its strength lies in the high reliability and precision of its expert-generated responses, though the queries are specifically tailored to the included mobile applications. Similarly, the Usable Privacy Policy Project~\cite{sadeh2013usable} employs machine learning and NLP to analyze and summarize privacy policies. As a result, the "OPP-115 Corpus" dataset~\cite{wilson2016creation} consists of 115 website privacy policies annotated with diverse information types.



\textbf{genAI Evaluation.} Recent research has significantly advanced our understanding of Large Language Models (LLMs). Ge et al.~\cite{ge2023openagi} demonstrated on the OpenAGI platform that domain-enhanced, optimized smaller LLMs can surpass larger models through Task Feedback Reinforcement Learning. Kang et al.~\cite{kang2023llms} explored LLMs in understanding user preferences, noting their competitive performance against traditional Collaborative Filtering methods when fine-tuned, despite initial shortcomings in zero-shot and few-shot scenarios. Chiang and Lee~\cite{chiang2023large} found a strong correlation between LLM and human evaluations in text quality assessments, especially with advanced models like InstructGPT and ChatGPT. Liu et al.~\cite{liu2023agentbench} introduced AgentBench, a benchmark focusing on LLMs as decision-making agents in interactive settings. Bang et al.~\cite{bang2023multitask} examined ChatGPT across various tasks, highlighting its limitations in low-resource and non-Latin languages. Zhang et al.~\cite{zhang2023benchmarking} cautioned about the potential inaccuracies in LLM-generated news summaries. Finally, Liu et al.~\cite{liu2023code} used EvalPlus to reveal previously unnoticed errors in LLM-generated code, emphasizing the need for robust evaluation. Collectively, these studies highlight the importance of diverse and comprehensive metrics for the effective and safe deployment of LLMs.

\textbf{General Question-answering Benchmarks.} Question-answering benchmarks like SQuAD~\cite{rajpurkar2016squad}, TriviaQA~\cite{joshi2017triviaqa}, and Holistic Evaluation of Language Models (HELM)~\cite{liang2022holistic} are pivotal in evaluating LLMs. These benchmarks consist of diverse questions, ranging from factual to complex reasoning tasks, and are typically derived from domains like Wikipedia or news articles. To gauge LLM performance, they employ metrics such as accuracy, precision, recall, and F1 score, focusing on answer quality aspects like clarity, relevance, and completeness. The HELM initiative stands out for its multi-metric approach and extensive evaluation across various language models, scenarios, and metrics, aiming for a thorough understanding of these models' capabilities, limitations, and potential risks. TriviaQA introduces a unique challenge by offering over 650,000 question-answer pairs covering a broad spectrum of topics, from science to popular culture. Its distinctiveness lies in its requirement for systems to retrieve and integrate information from diverse sources, as it presents questions independent of specific contexts.

\section{The \benchmark Benchmark}
\label{sect:benchmark}

\begin{figure*}[!htb]
    \centering
    \includegraphics[width=\textwidth]{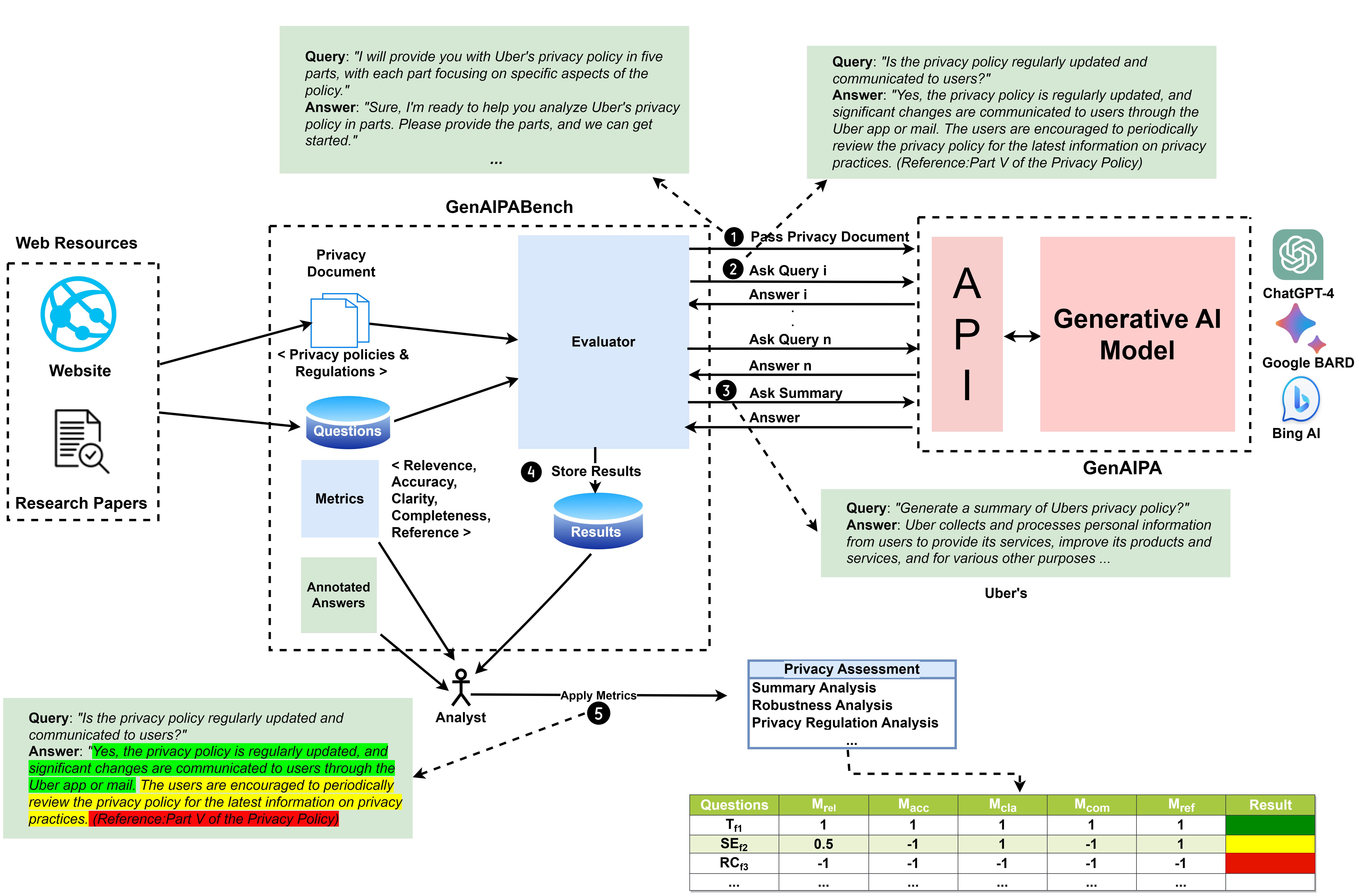}
    \caption{A high-level overview of \benchmark.}
    \label{fig:highlevel}
\end{figure*}

The \benchmark benchmark assesses generative AI-based privacy assistants (GenAIPAs), focusing on their ability to aid users in understanding the intricate realm of data privacy, namely: 1) Answering questions an individual might have about the privacy policy of an organization/corporation/service; 2) Answering questions about data privacy regulations in a specific country/state; 3) Summarizing privacy policies and privacy regulations. \benchmark comprises privacy documents, questions (with variations), annotated answers, and an evaluation tool (see Figure~\ref{fig:highlevel}). The full benchmark, as well as the results obtained evaluating three popular genAI systems (see Section~\ref{sect:results}) has been made available online\footnote{\url{https://anonymous.4open.science/r/GenAIPABench-FAB5/}}. 

\textbf{Privacy documents:} Extracted from web resources, the current version of \benchmark includes five privacy policies and two data regulations with their corresponding manually annotated answers to questions. This dataset equips GenAIPAs with specific content knowledge, facilitating a uniform comparison across various models, regardless of their prior training on these documents.

\textbf{Privacy questions:} Intended to test GenAIPAs' proficiency in interpreting and responding to typical queries about website/service privacy policies and regulations. The dataset contains 32 questions for privacy policies and six questions for privacy regulations covering crucial topics like data collection, storage, sharing, and user rights (see Section~\ref{sect:questions}). Along with the questions, the benchmark includes a set of paraphrased questions and variations for each.

\textbf{Metrics:} These criteria assess GenAIPAs' effectiveness in answering privacy policy and regulation questions. Metrics include accuracy, relevance, clarity, completeness, and reference, as detailed in Section~\ref{sect:metrics}. Human analysts use these metrics to review the responses generated by GenAIPAs and pinpoint areas needing enhancement.
    
\textbf{Annotated answers:} For the five privacy policies and two regulations included in the corpus, we meticulously curated answers for each benchmark question. This process involved two experts, each responsible for a different privacy policy, who created answers based on their assigned documents. After the initial answer generation, they conducted a reciprocal review, cross-verifying the responses against the original policies and refining them as needed. This rigorous process guarantees the precision and thoroughness of the annotated answers.

\textbf{Evaluator:} The evaluator automates the generation of prompts to introduce GenAIPAs to the privacy documents and pose the benchmark questions (see Appendix~\ref{sect:evaluator}). If an API is available, it also executes the prompts and handles the collection of answers. The evaluator initializes the GenAIPA with a prompt that includes information about which privacy document to refer to. The evaluator uses three different types of initialization prompts:

\begin{enumerate}
    \vspace{-0.2cm}\item Benchmark execution without accompanying privacy policy document: The evaluator prompts the GenAIPA, explaining that it will ask questions about a specific privacy document (e.g., the privacy policy of Uber).

    \item Benchmark execution with accompanying privacy document: The initial prompt explains that the evaluator will send the privacy document in segmented portions due to possible token limit constraints of GenAIPAs, followed by questions about the privacy document.

    \item Benchmark execution on summarized privacy document: The initial prompt requests the GenAIPA to summarize the privacy document (both with and without explicit privacy document introduction).

\end{enumerate}

The benchmark questions are posed following this prompt. The conversation is reset before performing the next type of  initialization prompt. 
The process is repeated multiple times (the number of repetitions is configurable), and a conversation reset is forced after each repetition.

\section{Question Corpus}
\label{sect:questions}
We introduce the question corpus that represents privacy questions an individual might ask the GenAIPA.

\subsection{Privacy Policy Questions}

To evaluate GenAIPA's performance comprehensively, we gathered questions spanning a broad spectrum of privacy-related topics concerning organizational or service privacy policies. These were grounded in established privacy frameworks and guidelines, as well as web resources. Initially, we selected pertinent privacy categories from the existing literature, notably referencing the ISO/IEC 29100:2011 - Information Technology - Security Techniques - Privacy framework~\cite{noauthor_iso/iec_nodate}, which offers a detailed privacy management framework, encompassing guidelines for privacy impact assessments and policies~\cite{IPC,pollach2007,oaic2023}. For each of the eight categories, we identified four questions\footnote{We limited the number of questions per category to four because of the intensive manual effort needed to generate and validate ground truth answers for each policy-question combination.} per category, with two sets of questions: 1) General user concerns and 2) Questions from specific individuals.
For the first set, we scoured privacy FAQs on websites with common user questions about data management. We also searched online forums like Reddit and Twitter with keywords from the privacy categories. The questions were generalized and combined to extract three per category. For the second set, we incorporated questions from a user study by Abhilasha et al.~\cite{ravichander2019question} that used Amazon Mechanical Turk to collect lay user questions about various app policies. We selected one question per category. Note that some questions include the placeholder \textit{[the company]}, to be replaced by the evaluator with the relevant company name when generating prompts for the assistant.

The final corpus of questions includes not only the original queries but also their paraphrased variants (i.e., reworded versions of the questions that maintain their original meaning). We introduce these variants to evaluate GenAIPAs' understanding and response abilities across diverse linguistic scenarios reflecting, for instance, different privacy knowledge of individuals. We used two approaches to generate the variations. For questions about general user concerns, we used QuillBot\footnote{\url{https://www.quillbot.com}}, an AI tool that automatically restructures sentences and alters words or phrases while preserving their original intent. We automatically generated ten variations per question, ultimately selecting three that showed diversity while guaranteeing that the original meaning was preserved. As an example, for the original question, "Does [the company] minimize data retention periods?" we generated variants such as "Does [the company] keep data for shorter times?" and "Does [the company] hold onto data for less time?". For each individual-specific question, we manually select, from the same dataset, three questions from other individuals with the same meaning but different wording.

\noindent In the following, we introduce each category and its questions. We will denote the generalized and individual-specific questions with the subscripts $f_{1/2/3}$ and $u_{1}$, respectively. 

\textbf{Transparency ($T$)} refers to how easily users can understand and access information regarding the collection, usage, and sharing of their personal data by companies or organizations. Key elements include the types of data collected, its intended use, and any third-party sharing. Crucial to transparency is using clear, understandable language and the accessibility of privacy policies. To evaluate transparency in privacy policies, the following questions are proposed:

\begin{itemize}
    \item[$\mathbf {T_{f_1}}$] \textit{``Does the policy outline data collection practices?''}
    \item[$\mathbf{T_{f_2}}$] \textit{``What is [the company]'s stance on government requests for user data?''}
    \item[$\mathbf{T_{f_3}}$] \textit{``How does the policy address potential conflicts of interest in data usage or sharing?''}
    \item[$\mathbf{{T_{u_1}}}$]\textit{``What sort of data is collected from me while using this?''}
\end{itemize}

$\mathbf {T_{f_1}}$ is a straightforward yes or no question that does not require much explanation or context. $\mathbf{T_{f_2}}$ asks about the company's stance on government requests for user data, which may require some knowledge of privacy regulations and the company's policies. $\mathbf{T_{f_3}}$ addresses potential conflicts of interest in data usage and sharing, a more nuanced and complex issue requiring a deeper understanding of the company's business practices and policies. Finally, $\mathbf{{T_{u_1}}}$ informs users about the specific data types collected while interacting with the service or product.

\textbf{User Control ($UC$)} refers to the options available to users to manage their personal information and privacy settings. These controls can include the ability to opt out of data collection and sharing, to delete personal data, to access and modify personal data, and to set preferences for how their data is used. To evaluate user control in privacy policies, the following questions are proposed:

\begin{itemize}
    \item[$\mathbf{UC_{f_1}}$] \textit{``Are users given control over their data and privacy settings?''}
    \item[$\mathbf{UC_{f_2}}$] \textit{``Are there clear mechanisms for users to request data deletion or
    access?''}
    \item[$\mathbf{UC_{f_3}}$] \textit{``How does [the company] manage consent and withdrawal of consent from users?''}
    \item[$\mathbf{{UC_{u_1}}}$]\textit{``Can I opt out of letting them collect data and still use the app?''}
\end{itemize}

$\mathbf{UC_{f_1}}$ checks for the fundamental aspect of a privacy policy: whether it empowers users to manage their data. $\mathbf{UC_{f_2}}$ delves into the company's data deletion and access procedures, requiring detailed knowledge of their data management practices.$\mathbf{UC_{f_3}}$ explores the complexities of how the company navigates user consent and its revocation, an area influenced by both the company's specific policies and the legal framework it operates within. Finally, $\mathbf{{UC_{u_1}}}$  examines whether the policy allows users to decline data collection while maintaining access to the app, reflecting a critical aspect of user control and consent in privacy practices.

\textbf{Data Minimization and Purpose Limitation ($DM$)} are key principles safeguarding user privacy. Data minimization restricts the collection, use, and storage of personal data to essentials, mitigating risks and preventing misuse for unrelated purposes. Conversely, Purpose Limitation confines data use to its original collection intent, giving users more control over their information.
To evaluate data minimization and purpose limitation in privacy policies, the following questions are proposed:

\begin{itemize}
    \item[$\mathbf{DM_{f1}}$] \textit{``Does [the company] minimize data retention periods?''}
    \item[$\mathbf{DM_{f2}}$] \textit{``How is user data anonymized or aggregated to protect individual privacy?''}
    \item[$\mathbf{DM_{f3}}$] \textit{``Are there any restrictions on data processing for specific purposes or contexts?''}
    \item[$\mathbf{DM_{u1}}$] \textit{``How long is my data stored?''}
\end{itemize}

$\mathbf{DM_{f1}}$ inquires if the company minimizes data retention periods, seeking a direct response based on the company's data retention policy. $\mathbf{DM_{f2}}$ delves into the techniques for anonymizing or aggregating user data to safeguard privacy, which may demand technical insight for a comprehensive answer. $\mathbf{DM_{f3}}$ probes into any constraints on data processing tailored to particular purposes or contexts, necessitating a thorough understanding of the company's policies and legal obligations. Finally, $\mathbf{DM_{u1}}$ seeks specific information about the length of time the company retains user data.

\textbf{Security and Encryption ($SE$)}  involves strategies organizations use to safeguard users' personal data from unauthorized access, theft, or cyber-attacks. This includes employing encryption for sensitive data like usernames, passwords, and credit card details, and using secure communication protocols to avert data interception. Additionally, organizations often establish policies for managing security breaches, encompassing user notification, breach investigation, and preventive measures for future incidents. To assess security and encryption in privacy policies, consider these questions:

\begin{itemize}
    \item[$\mathbf{SE_{f_1}}$] \textit{``Are user communications encrypted end-to-end?''}
    \item[$\mathbf{SE_{f_2}}$] \textit{``What measures are in place to prevent unauthorized access to user data?''}
    \item[$\mathbf{SE_{f_3}}$] \textit{``How are data breaches or security incidents handled and communicated to users?''}
    \item[$\mathbf{SE_{u_1}}$] \textit{``How well secured is my private information?''}
\end{itemize}

$\mathbf{SE_{f_1}}$ checks if the company uses end-to-end encryption for user communications, requiring a simple yes or no answer. $\mathbf{SE_{f_2}}$ inquires about specific security measures against unauthorized data access, calling for a detailed response about the company's security protocols. $\mathbf{SE_{f_3}}$ explores the handling and communication of data breaches or security incidents, necessitating an understanding of the company's response strategy and relevant legal/regulatory frameworks. Finally, $\mathbf{SE_{u_1}}$ seeks a general assessment of the overall security measures in place to protect users' private information. 

\textbf{Privacy by Design and Innovation ($PbD$)}  embodies a data protection strategy that integrates privacy considerations throughout all stages of product or service development. This method involves embedding privacy-enhancing features, like data minimization, purpose limitation, and robust security measures, from the outset. The aim is to proactively mitigate privacy risks and ensure default data protection. Moreover, $PbD$ advocates continuous monitoring and updating of privacy practices to address emerging privacy concerns. To assess $PbD$ in privacy policies, consider these questions:

\begin{itemize}
    \item[$\mathbf{PbD_{f_1}}$] \textit{``Does [the company] conduct privacy impact assessments?''}
    \item[$\mathbf{PbD_{f_2}}$] \textit{``Are there privacy-enhancing technologies implemented, such as differential privacy?''}
    \item[$\mathbf{PbD_{f_3}}$] \textit{``Does [the company] use automated decision-making or profiling, and if so, how does it impact user privacy?''}
    \item[$\mathbf{PbD_{u_1}}$] \textit{``What sort of analytics will my data be subjected to?''}
\end{itemize}

$\mathbf{PbD_{f_1}}$ is a straightforward yes-or-no question about whether the company conducts privacy impact assessments, a standard procedure in data privacy. $\mathbf{PbD_{f_2}}$ involves the concept of differential privacy, a more advanced and technical area that requires a nuanced understanding of how to balance data utility and privacy. $\mathbf{PbD_{f_3}}$ is about the complex topics of automated decision-making and profiling, which demand a deep technical understanding and the ability to assess ethical and privacy implications. Finally, $\mathbf{PbD_{u_1}}$  seeks to understand how the company uses, analyzes, and potentially benefits from user data while also considering the privacy implications of such analytics.

\textbf{Responsiveness and Communication ($RC$)} pertains to how organizations interact with users regarding privacy matters. This encompasses providing transparent, easily understandable information on data practices and swiftly addressing user privacy queries and concerns. To assess these aspects in privacy policies, consider the following questions:

\begin{itemize}
    \item[$\mathbf{RC_{f_1}}$] \textit{``Is the privacy policy regularly updated and communicated to users?''}
    \item[$\mathbf{RC_{f_2}}$] \textit{``Is there a process in place to address user privacy complaints?''}
    \item[$\mathbf{RC_{f_3}}$] \textit{``Does [the company] publish transparency reports detailing government data requests, surveillance, or law enforcement interactions?''}
    \item[$\mathbf{RC_{u_1}}$] \textit{``Has there ever been a security breach?''}
\end{itemize}

$\mathbf{RC_{f_1}}$ is straightforward, seeking a yes or no answer regarding the communication of privacy policy updates. $\mathbf{RC_{f_2}}$ delves into the company's mechanisms for handling privacy complaints, requiring an understanding of their specific procedures. $\mathbf{RC_{f_3}}$, more intricate, probes into the company's transparency regarding governmental data requests and legal interactions, demanding insight into their commitment to transparency and legal compliance. Finally, $\mathbf{RC_{u_1}}$ is highly relevant in the context of how a company manages and communicates about security incidents, a critical aspect of user trust and data protection.

\textbf{Accessibility, Education, and Empowerment ($AEE$)} focuses on ensuring that privacy policies are user-friendly and empowering. Policies should be accessible, including to those with disabilities, through various formats like audio or video. They need to be in plain language for easy comprehension, explaining key concepts and terms clearly. It is crucial to educate users about their privacy rights and the implications of data sharing. Policies should guide users on how to exercise their privacy rights and control their personal data. Empowerment is key, providing users with meaningful choices in a straightforward manner. The following questions are proposed to evaluate these aspects of privacy policies:

\begin{itemize}
    \item[$\mathbf{AEE_{f_1}}$] \textit{``Are employees trained on data privacy best practices and handling sensitive information?''}
    \item[$\mathbf{AEE_{f_2}}$] \textit{``How are user data privacy preferences managed across different devices or platforms?''}
    \item[$\mathbf{AEE_{f_3}}$] \textit{``Does [the company] offer user-friendly resources, such as tutorials or guides, to help users effectively manage their privacy settings and understand their data rights?''}
  \item[$\mathbf{AEE_{u_1}}$] \textit{``Does it share any data with a third party?''}
\end{itemize}

$\mathbf{AEE_{f_1}}$ is straightforward, asking whether the company ensures its employees are trained in data privacy. $\mathbf{AEE_{f_2}}$ inquires about managing user privacy across various platforms, requiring an understanding of integrated privacy systems. $\mathbf{AEE_{f_3}}$ delves into the availability of educational resources for users, indicating the company’s commitment to user education in privacy. $\mathbf{AEE_{u_1}}$ directly addresses the transparency of the company's data-sharing policies, which is a fundamental aspect of user trust and privacy management.

\textbf{Compliance and Accountability ($CA$)} are critical for organizations to ensure adherence to privacy laws and standards. This includes conducting regular audits, performing data protection impact assessments, and appointing a Data Protection Officer (DPO) to oversee privacy matters. Accountability extends to taking responsibility for privacy breaches or violations and providing remedies to affected parties. The proposed questions for evaluating compliance and accountability in privacy policies are:

\begin{itemize}
    \item[$\mathbf{CA_{f_1}}$] \textit{``Does the policy comply with applicable privacy laws and regulations?''}
    \item[$\mathbf{CA_{f_2}}$] \textit{``What steps are taken to ensure data processors and subprocessors adhere to privacy requirements?''}
    \item[$\mathbf{CA_{f_3}}$] \textit{``Does [the company] have a process in place for reporting and addressing privacy violations or non-compliance issues, both internally and with third-party vendors?''}
    \item[$\mathbf{CA_{u_1}}$] \textit{``Do I have any rights as far as whether I want my account info deleted?''}
\end{itemize}

$\mathbf{CA_{f_1}}$ assesses the company's alignment with privacy laws, a fundamental aspect of privacy management. $\mathbf{CA_{f_2}}$ explores how the company ensures that it's data processors and subprocessors comply with privacy standards, reflecting an advanced understanding of third-party risk management. $\mathbf{CA_{f_3}}$ inquires about the procedures for handling privacy violations, demonstrating the depth of the company's commitment to accountability. Finally, $\mathbf{CA_{u_1}}$ directly relates to the "right to be forgotten," a key provision of regulations such as GDPR, which empowers individuals to request the deletion of their personal data under certain circumstances.

\subsection{Privacy Regulation Questions}

\benchmark includes questions to evaluate the performance of the GenAIPA in helping users understand privacy and data protection regulations such as the GDPR or the CCPA. We compiled and generalized the following questions extracted from different sources~\cite{gdpr_faq,ccpa_faq} that aim to cover a range of topics, from the scope and applicability of the regulations to specific requirements and rights:

\begin{enumerate}
    \item[$\mathbf{PR_1}$] \textit{``Who must comply with the [regulation]?''}
    \item[$\mathbf{PR_2}$] \textit{``What are the [regulation] fines?''}
    \item[$\mathbf{PR_3}$] \textit{``How do I comply with the [regulation]?''}
    \item[$\mathbf{PR_4}$] \textit{``Does the [regulation] require encryption?''}
    \item[$\mathbf{PR_5}$] \textit{``What is personal information and sensitive personal information under the [regulation]?''} 
    \item[$\mathbf{PR_{6}}$] \textit{``What rights do I have under the [regulation]?''}
\end{enumerate}

Note that the evaluator will replace the placeholder \textit{[regulation]} with the specific privacy regulation to be evaluated from the privacy document dataset (e.g., GDPR, CCPA, LGPD, etc.). Like the prior question corpus, the benchmark includes question variations through paraphrasing for comprehensive evaluation.

\section{Metrics}
\label{sect:metrics}
To assess the GenAIPA's response quality, we developed metrics integrating five principal elements anchored in privacy policy assessment. These metrics draw on resources like the Future of Privacy Forum's report and other key studies~\cite{future_of_privacy_forum_best_2020, martin2019,Bamberger2011PrivacyOT} which offer valuable guidance on designing and evaluating privacy policies.

\textbf{Relevance ($\mathcal{M}_{rel}$)} gauges the alignment of an answer with the user's question, a critical factor for user satisfaction in conversational agents~\cite{bickmore2018maintaining}. Relevant responses empower users to make knowledgeable decisions about their data privacy whereas irrelevant answers may cause frustration and dissatisfaction, obstructing users' comprehension of their rights and responsibilities~\cite{luger_sellen_2016}.

\textbf{Accuracy ($\mathcal{M}_{acc}$)} evaluates the correctness of information provided by AI systems, crucial for fostering trust and acceptance, as emphasized in~\cite{liao_gao_wu_zhang_2019}. Inaccurate or misleading information can lead to poor decisions, adversely affecting user perception of the system. Specifically, incorrect responses by a GenAIPA can result in misguided actions regarding privacy, such as unwisely continuing to use a service perceived as less intrusive. Moreover, recognizing inaccuracies can diminish the perceived reliability and trustworthiness of the system, impacting user confidence~\cite{CHOI201842}.  


\textbf{Clarity ($\mathcal{M}_{cla}$)} assesses the effectiveness of communication, focusing on clear and coherent responses, as per Grice's principles~\cite{grice_1975}. It emphasizes the importance of easily understandable and coherent responses for informed decision-making. A major challenge with privacy policies, as noted in~\cite{jensen2004privacy}, is their complexity due to legal and technical jargon. GenAIPAs should strive for simplicity, avoid ambiguity and unnecessary technical terms, and provide clear explanations. Tailoring responses to the user's comprehension level is also vital. By ensuring clarity, GenAIPAs improve user satisfaction and guarantee effective information transmission.

\textbf{Completeness ($\mathcal{M}_{com}$)} measures if an answer fully addresses the user's question~\cite{radziwill_benton_2017}. Responses must encompass all necessary aspects and details, avoiding the need for multiple follow-up questions. A complete answer should thoroughly cover the topic, provide accurate and exhaustive information, and consider any related issues pertinent to the user's query. Inadequate or flawed information can lead to misinformed decisions or a lack of understanding regarding privacy options, resulting in user frustration and diminished trust in the AI system~\cite{CHOI201842}. To ensure completeness, GenAIPAs must understand the context of queries and tailor responses to meet specific user needs. This approach not only boosts user satisfaction but also streamlines communication.

\textbf{Reference ($\mathcal{M}_{ref}$)} evaluates the inclusion of proper citations or mentions of relevant policy sections in responses, crucial for transparency and credibility in legal or policy contexts, as underscored in~\cite{savelka_ashley_2016}. AI systems, when applicable, should incorporate accurate citations or references to pertinent policy sections. This practice bolsters the response's accuracy and completeness and enhances user trust by providing transparency and credibility. Proper citation entails using the correct legal or policy language, including relevant section numbers and any other information essential for comprehending the legal or policy implications of the user's query. By integrating appropriate references in their responses, GenAIPAs can assure users of the accuracy and compliance of their responses with relevant laws or policies.

\paragraph{Metric Evaluation.} The proposed evaluation method assesses each response across five metrics on a scale from +1 to -1:

\begin{itemize}
    \item $\mathcal{M}_{rel}$: +1 for a relevant response, +0.5 for a partially relevant response, and -1 for a not relevant response.
    \item $\mathcal{M}_{acc}$: +1 for an entirely correct response, +0.5 for a partially correct response, and -1 for an incorrect response.
    \item $\mathcal{M}_{cla}$: +1 for a clear and easy-to-understand response, +0.5 for a somewhat clear but could be improved response, and -1 for a confusing or hard-to-comprehend response.
    \item $\mathcal{M}_{com}$: +1 for a comprehensive response, +0.5 for a somewhat complete but lacking some minor information response, and -1 for an incomplete or missing important details response.
    \item $\mathcal{M}_{ref}$: +1 for a correctly cited relevant policy section, +0.5 for a mentioned section without explicitly citing it, and -1 for an incorrect reference.
\end{itemize}

Note that it is possible that a specific privacy document (e.g., a specific privacy policy) might lack information to answer a benchmark question. In that case, the desired answer should be that the document does not contain enough information to answer the question. Hence, any mention to a policy section would score a -1 for $\mathcal{M}_{ref}$.
We propose to aggregate these into an overall quality metric ($\mathcal{M}{all}$) by calculating total positive/partial points ($\mathcal{M}_{all}^{+}$) and total negative points ($\mathcal{M}_{all}^{-}$) separately. The overall score is normalized using the equation:
\begin{equation*}
        \mathcal{M}_{all} = \frac{(\text{Current Score} - \text{Minimum Score})}{(\text{Maximum Score} - \text{Minimum Score})} \times 9 + 1
\end{equation*}
\noindent where the Minimum Score is -5 and the Maximum Score is 5. This approach highlights the potential negative impact of answers on privacy decision-making.


\section{Experiments}
\label{sect:results}
We assessed three leading generative AI systems using \benchmark: ChatGPT-4~\cite{openai2023gpt4}, Bard~\footnote{\url{https://bard.google.com/}}, and BingAI\footnote{\url{https://chat.openai.com/}}. ChatGPT-4, accessed via OpenAI's API\footnote{\url{https://platform.openai.com/}}, and Bing AI, both based on GPT-4, differ in their fine-tuning and deployment, influencing their functionalities and user interactions. Bard and BingAI were accessed through their websites, as no official APIs were available. Our evaluation analyzed five privacy policies (Uber, Spotify, Airbnb, Facebook, Twitter) and two major privacy regulations (GDPR, CCPA). We included statistics about the selected privacy policies in Table \ref{table:policy_analysis}. The policy's unique word frequency \cite{Hiebert} indicates whether complex and/or specialized language is used, which might challenge LLMs if it is beyond their training. The estimated reading time (computed as the number of words multiplied by the average time in minutes required per word) represents the length of the document, which might impact response coherence and relevance. The reading level (Flesch-Kincaid Grade Level \cite{Flesch2007}) metric assesses text difficulty, indicating the education level needed for understanding. Finally, although long, connective-word-rich sentences can confuse humans, they might help GenAIPAs understand context and logic. Additionally, we note that out of 32 questions, 15 (Facebook), 14 (Twitter), 13 (Airbnb), and 8 (Spotify, Uber), were  on content not explicitly stated in the policy.


\begin{table}[!htb]
\centering
\caption{Privacy policies analyzed.}
\label{table:policy_analysis}
\begin{tabular}{|p{1.2cm}|p{1cm}|p{1cm}|p{1cm}|p{1.4cm}|}
\hline
Policy & Unique Words & Reading Time & Reading Level & Connective Words \\
\hline
Twitter & 0.21 & 21 & 10.3 & 0.04\\
Spotify & 0.16 & 32 & 12.4 & 0.03\\
Uber & 0.16 & 37 & 11.9 & 0.04\\
Airbnb & 0.19 & 27 & 14.1 & 0.04\\
Facebook & 0.18 & 20 & 11.8 & 0.05\\
\hline
\end{tabular}
\end{table}

The following sections present the analysis of the results obtained. The performance results are plotted in graphs (e.g., Figure~\ref{fig:iqr_orig}) that show the performance score, calculated using interquartile range (IQR) values from questions. To enhance visual interpretation, average performance scores are also mapped across five metrics using heatmaps (e.g., Figure~\ref{fig:original_combined}), where the x-axis represents the chosen metrics and the y-axis corresponds to the different categories of privacy questions).

\subsection{Assessing the Quality of Responses to Privacy Policy Questions}

\begin{figure*}[!htb]
     \centering
     \begin{subfigure}[b]{0.80\textwidth}
         \centering
         \includegraphics[width=\textwidth]{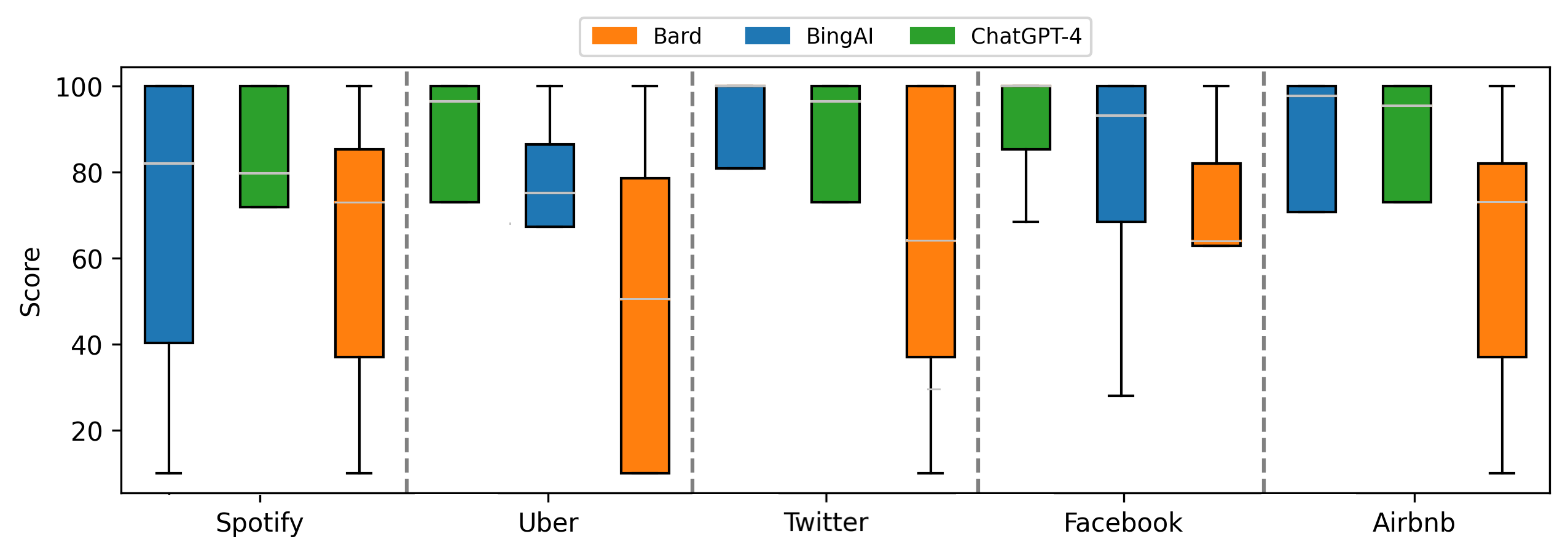}
          \caption{Overall score per policy.}
         \label{fig:iqr_orig}
     \end{subfigure}
     \begin{subfigure}[b]{0.99\textwidth}
         \centering
         \includegraphics[width=\textwidth]{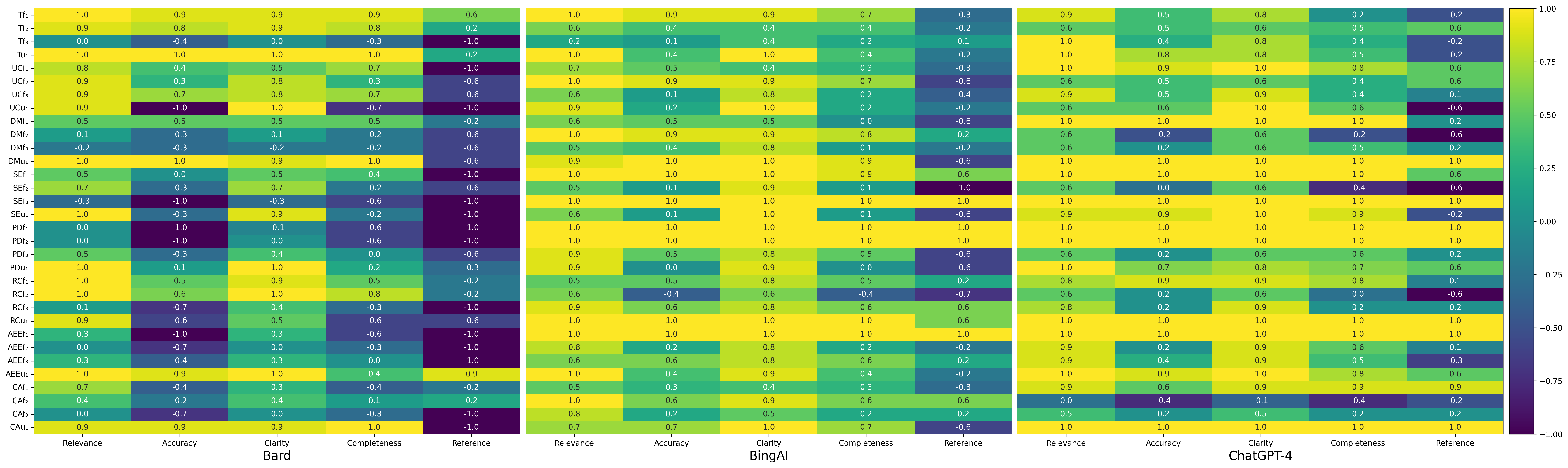}
          \caption{Average scores for all policies across metrics.}
          \label{fig:original_combined}
     \end{subfigure}
     \caption{Performance of systems when the privacy policy is explicitly shared.}
\end{figure*}

This experiment aims to assess the quality of responses concerning privacy policy questions. 
The results (see Figure~\ref{fig:original_combined}) show that ChatGPT-4 and BingAI consistently outperform Bard in most questions. Notably, BingAI stands out in its ability to adeptly handle user and FAQ-generated questions, especially in the context of Spotify, Twitter, and Airbnb policies. This proficiency may be due to a simpler reading level, a more diverse vocabulary, and lower reading times of the policies (see Table \ref{table:policy_analysis}). Bard's performance tends to diminish as question complexity increases, a trend not observed as prominently in ChatGPT-4 or BingAI. We next analyze in detail the performance of each system.

\textbf{ChatGPT-4}: While ChatGPT-4 often achieved a median score of 100 (see Figure \ref{fig:iqr_orig}), its performance varied significantly with scores ranging from 10 to 100 across all policies. This fluctuation was particularly noticeable in responses to FAQ-sourced questions. Its performance dipped in handling questions on the Spotify policy, with a lower median score of 79.75, compared to the Facebook and Uber policies, which had median scores of 100 and 96.5, respectively. The interquartile range further illustrated this trend. ChatGPT-4's relevance (Figure \ref{fig:original_combined}) in answering questions was generally strong, with scores ranging between 0.6 to 1 for most categories. However, it seemed to struggle slightly with $CA_{f2}$ at 0. The clarity exhibited by ChatGPT-4 was commendable, consistently hovering between 0.6 and 1, except for a noticeable dip to -0.1 for $CA_{f2}$. Accuracy, however, was a mixed bag, while ChatGPT-4 scored admirably with a peak of 1 for $SE_{f1}$, it descended to -0.4 for $CA_{f2}$. Completeness followed a similar trajectory, ranging from highs like 1 ($SE_{f3}$) to lows of -0.4 ($CA{f2}$). ChatGPT-4’s referencing capabilities appeared as an area of improvement, with several scores lying in the negative domain. Moreover, ChatGPT-4 showed consistently strong performance across all policies, without a specific trend towards those with higher or lower proportions of non-existent content.

\textbf{Bard}: Bard (see Figure \ref{fig:iqr_orig}) frequently registered minimum scores of 10 across various question categories and very occasionally scored higher (it peaked at around 100 for the combination of the Spotify policy and user-generated questions). The median scores provide further insights into its tendency to gravitate towards mid-range values for all questions, evidenced by median scores like 50.5 (Uber) and 64 (Twitter). User-generated questions yielded higher median scores than FAQ questions, with the Bard policy excelling in Spotify-related questions at a median of 95.5. Additionally, the Twitter policy outperformed others, with a median score of 64.4 for FAQ questions. Bard's 1st quartile performance for questions often struggled, while its 3rd quartile results indicated that even its top performance strata seldom achieved peak scores. Figure \ref{fig:original_combined} shows that Bard's relevance was high across the board, with scores largely hovering around 1, though it faced challenges with $SE_{f3}$ at -0.3. The clarity metric also displayed consistency, mostly remaining close to 0.9, but $SE_{f3}$ presented a deviation with a score of -0.3. Accuracy for Bard varied considerably: it showed robustness in questions like $T_{u1}$ and $DM_{u1}$ with scores at 1 but dipped to -1 for metrics like $SE_{f2}$ and $PD_{f1}$. Regarding completeness, Bard oscillated between a high of 1 ($T_{u1}$) to a low of -0.7 ($UC_{u1}$). The reference domain was particularly an issue for Bard, with scores mainly revolving around -1 across all the questions showcasing the least favourable outcome. Additionally, Bard's performance was consistently moderate across a range of policies, with a noticeable dip in performance when dealing with Uber's policy, which had less non-existent content but required more reading time. In contrast, Bard excelled with Facebook's policy, characterized by less reading time and a higher proportion of non-existent content.

\textbf{BingAI}: Of the three systems, BingAI consistently demonstrated superior performance metrics (Figure \ref{fig:iqr_orig}). Its score spectrum was high, frequently attaining the maximum score of 100 across various difficulties and policies, seldom dropping below 20 for a specific case (Uber, FAQ-sourced question). This performance was equally evident in the median values, where BingAI displayed high consistency even for all questions. Noteworthy were scores of 100 (Twitter, FAQ questions) and 95.5 (Airbnb, user-generated questions). FAQ-sourced questions yielded lower median scores than user-generated questions, with BingAI excelling in Airbnb-related privacy questions at a median of 100. Additionally, in the FAQ group, the Twitter policy outperformed others with a median score of 100 for user-generated questions. The quartile analysis reinforced its robustness, with the 1st quartile values indicating high baseline performance and the 3rd quartile metrics often culminating near or at 100. BingAI's performance showed consistently high values in several metrics (Figure \ref{fig:original_combined}). Its relevance and clarity stood out, surpassing the 0.8 mark. However, $DM_{f3}$ was an outlier in relevance with a score of -0.2. BingAI's accuracy demonstrated consistent strength, frequently achieving a score of 1, though significant challenges were noted in $DM_{f3}$ and $SE_{u1}$ with scores of -1 and 0, respectively. Regarding completeness, BingAI's metrics were predominantly positive, with a substantial number of questions securing a score of 0.9 or higher, but a noticeable decline was observed in $DM_{f3}$ at -1. Referencing for BingAI showed variance but managed to avoid deeply negative scores. BingAI consistently outperformed across different policies and questions (even for those questions on content not explicitly stated in the policy). Notably, a slight decline in BingAI's effectiveness was observed with Spotify's policy, characterized by lower non-existent content yet demanding more reading time. Conversely, BingAI's performance excelled with Twitter's policy, which required less reading time and contained a higher amount of non-existent content.

\subsection{Assessing Robustness through Paraphrased Questions}

\begin{figure*}[!htb]
     \centering
     \begin{subfigure}[b]{0.80\textwidth}
         \centering
         \includegraphics[width=\textwidth]{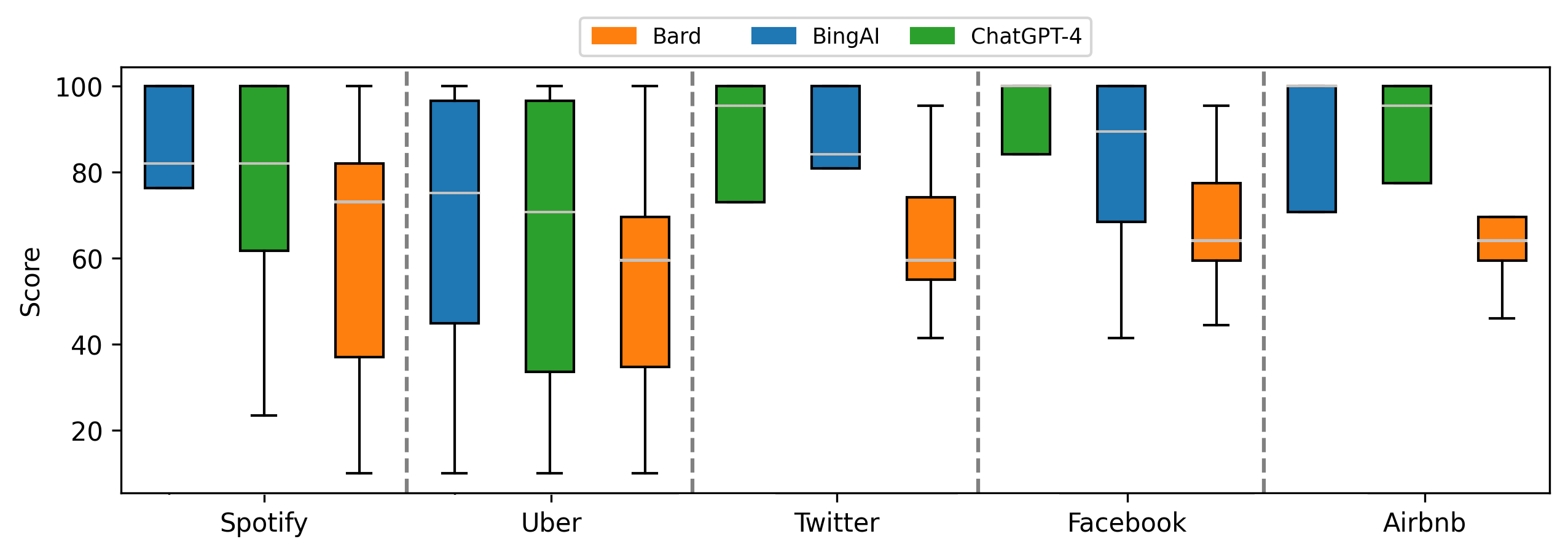}
          \caption{Overall score per policy.}
         \label{fig:iqr_robust}
     \end{subfigure}
     \hfill
     \begin{subfigure}[b]{0.99\textwidth}
         \centering
         \includegraphics[width=\textwidth]{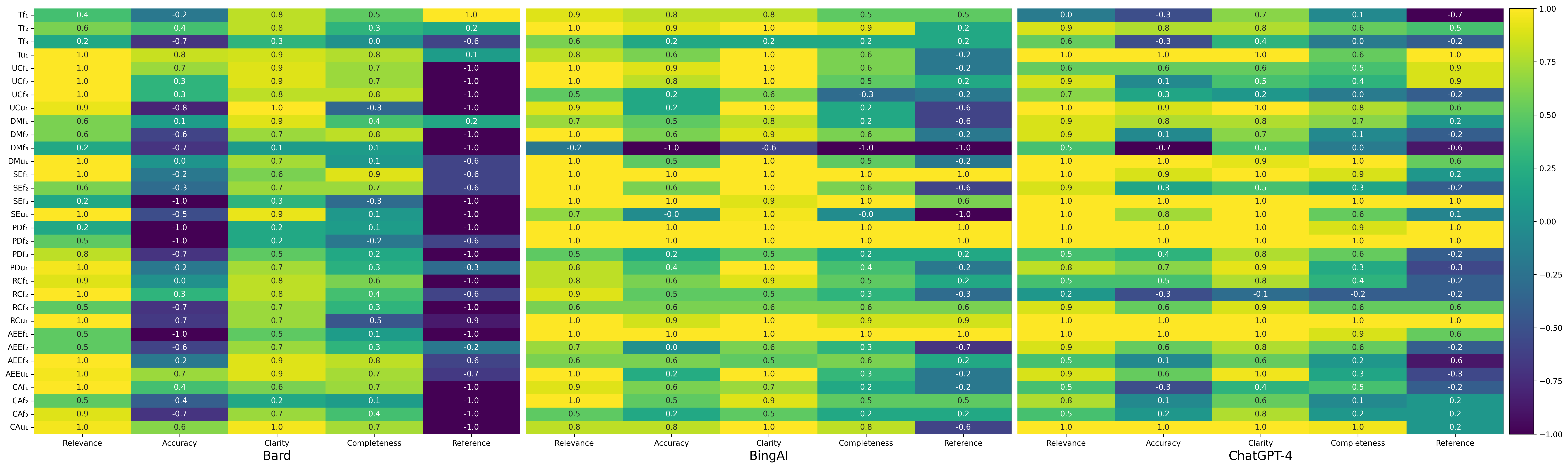}
          \caption{Average scores for all policies across metrics.}
         \label{fig:robust_combined}
     \end{subfigure}
     \caption{Performance of systems for paraphrased questions when the privacy policy is explicitly shared.}
     
\end{figure*}

The main goal of this experiment is to evaluate the robustness and consistency of the systems in providing similar responses to paraphrased variants of the questions. The results (see Figure~\ref{fig:iqr_robust}) show that ChatGPT-4 displayed consistent strengths, BingAI excelled in certain areas but showed referencing challenges, and Bard presented a mix of highs and noticeable lows. 

\textbf{ChatGPT-4}: ChatGPT-4 exhibited consistent performance across most policies (Figure \ref{fig:iqr_robust}), irrespective of questions coming from user or FAQ's. With Spotify, there was a decline in performance as we moved from FAQ questions to user-generated questions, from a median score of 86.5  to 76.3. Interestingly, the third quartile score remained around 100 for all questions, indicating that while the central tendency was lower, a subset of responses still reached the top performance. Twitter and Facebook cases showcased strong performance, with median scores not dipping below 70 across all questions. For Airbnb,  ChatGPT-4 answered with high proficiency for FAQ and user-generated questions, with the system achieving medians of 95.1 and 91, respectively. For Relevance, scores ranged between 0 and 1, showing high consistency in areas such as  $SE_{f1}$, $SE_{f3}$, $PD_{f1}$, and $PD_{f2}$ among others (Figure \ref{fig:robust_combined}). Clarity ratings showed a similar tendency, with the model performing excellently on queries like $SE_{f1}$ and $UC_{u1}$, scoring a perfect 1 while encountering challenges in $RC_{f2}$ and $UC_{f2}$. Accuracy results were more variable, with instances like  $DM_{u1}$, $SE_{f3}$, and $AEE_{f1}$' scoring near or at the top, juxtaposed against scores as low as -0.7 in $DM_{f3}$. Completeness spanned from high performances in $SE_{f3}$ to lows in $RC_{f2}$, $UC_{f3}$, and $DM_{f3}$. Reference scores showed strong points, such as 0.9 in $UC_{f1}$ and $UC_{f2}$, but also revealed potential areas of improvement with scores like -0.7 in $T_{f1}$. Additionally, ChatGPT-4 demonstrated strong performance across policies, particularly excelling in policies like Facebook's, which had a higher number of questions on content non-explicitly stated in the policy and required shorter reading times, achieving a median score near 100. However, there was a significant drop in performance with Uber's policy characterized by longer reading duration.

\textbf{Bard}: Bard's performance varied across policies (Figure \ref{fig:iqr_robust}). While FAQs and user-generated achieved median scores of 60.4 and 66.85, respectively, for Spotify, a significant drop to 59.5 was observed for user-generated questions. The minimum scores for the $SE_{f3}$,$PD{f2}$ among others were as low as 38.8 and 41.5, respectively, indicating challenging queries for Bard. Uber-related privacy questions posed difficulties across both categories, with the FAQ-sourced questions having a median of 61.75 but a minimum score of 1. Both Twitter and Facebook had mid-range median scores, with user-sourced questions in Facebook policy yielding a consistent median and third quartile, at 6 and 7.75, respectively. Airbnb responses were relatively stable, with scores fluctuating between 46 to 70. Analyzing Bard's performance across metrics (see Figure \ref{fig:robust_combined}), we observe that relevance ranged from scores as high as 1 for $UC_{f1}$, $UC_{f2}$, and $UC_{f3}$ to as low as 0.2 for $T_{f3}$ and $PD_{f1}$. Clarity was similarly distributed, with certain questions like $UC_{u1}$ receiving high scores of 1, while others, such as $DM_{f3}$, only achieved a score of 0.1. Accuracy proved to be a challenging area, with the lowest score being -1 for several questions, including $SE_{f3}$, $PD_{f1}$, and $AEE_{f1}$. However, the model managed to score 0.8 for $T_{u1}$. Completeness ranged from a notable 0.9 for $SE_{f1}$ to less promising results like -0.3 for $SE_{f2}$. The Reference metric had its highs and lows, with the highest score being 1 for $T_{f1}$ and several instances of -1, indicating an inconsistency in this domain. Finally, Bard also mirrored the ChatGPT performance on questions on content not explicitly stated in the policy, showing good performance, especially on Facebook's policy, while a performance reduction was noted with Uber's policy (the longest of the policies used).

\textbf{BingAI}: BingAI exhibited a mix of outstanding and lacklustre performances (Figure \ref{fig:iqr_robust}). For Airbnb, it achieved perfect medians of 100 for all questions, but the range was wide, from 10 to 100. The Uber policy was challenging, especially in user-generated questions, with a median of just 68.5 and a narrow range, indicating a uniform struggle. Twitter and Facebook policies saw robust results, with medians consistently above 84.5. For Airbnb questions, BingAI's performance was notable, particularly Twitter and Facebook in FAQ-sourced questions, where the system reached an almost perfect median score of 93.25.
BingAI demonstrated great performance in Relevance, particularly for questions like $T_{f2}$, $PD_{f2}$, $UC_{f1}$, and $SE_{f1}$, all scoring a perfect 1, but also showed weaker areas with scores like -0.2 for $DM_{f3}$ (Figure \ref{fig:robust_combined}). Clarity maintained a consistent trend, with scores predominantly leaning toward the higher end. For Accuracy, BingAI had top-performing scores in areas like $AEE_{f1}$, $PD_{f1}$, and $PD_{f2}$, but faltered in others, achieving a score of -1 for $DM_{f3}$. In terms of Completeness, it exhibited excellence in $SE_{f1}$ and $SE_{f3}$, both scoring 1, but saw a drop in areas like $DM_{f3}$. The Reference scores varied, ranging from 1 in $PD_{f1}$ and $PD_{f2}$ to lows of -1 in areas such as $DM_{f3}$. Additionally, BingAI consistently showed strong performance across all policies. However, there was a slight dip in its performance for Uber's policy like for other systems.

\subsection{Assessing the Ability to Recall Learned Privacy Policy Knowledge}

\begin{figure*}[!htb]
     \centering
     \begin{subfigure}[b]{0.90\textwidth}
         \centering
         \includegraphics[width=\textwidth]{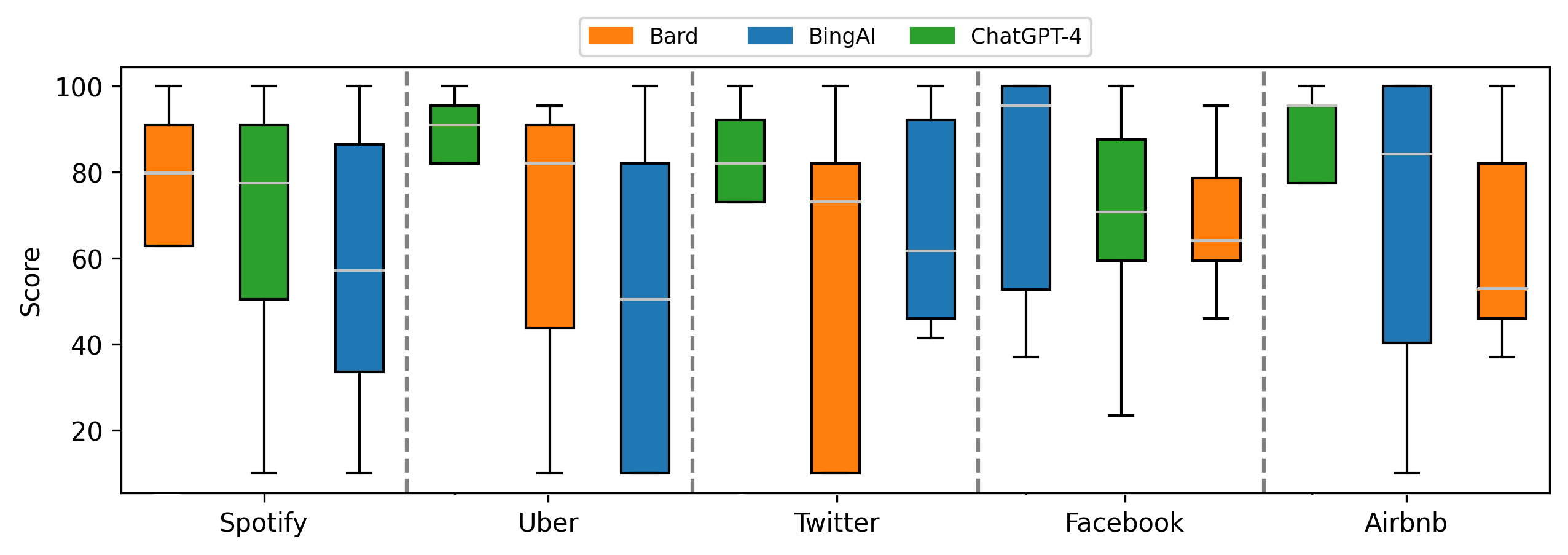}
          \caption{Overall score per policy.}
         \label{fig:iqr_adap}
     \end{subfigure}
     \hfill
     \begin{subfigure}[b]{0.99\textwidth}
         \centering
         \includegraphics[width=\textwidth]{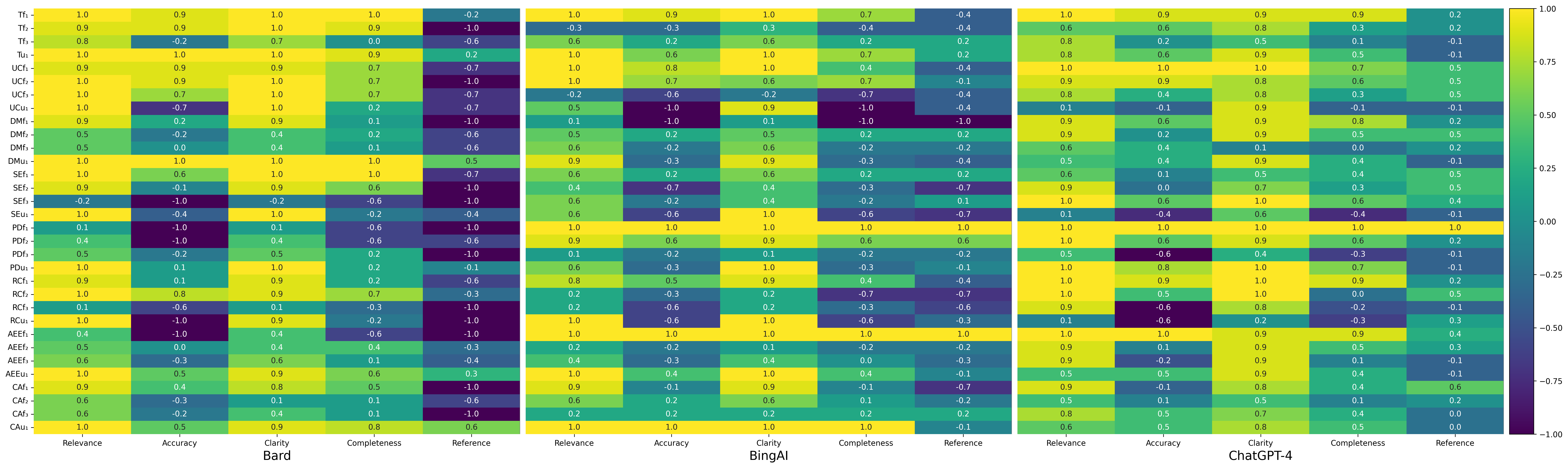}
          \caption{Average scores for all policies across metrics.}
         \label{fig:adaptive_combined}
     \end{subfigure}
     \caption{Performance of systems when the policy is not explicitly shared.}
\end{figure*}



The purpose of this experiment is to assess the performance of the systems when the privacy policy is not given explicitly. Hence, the system has to rely on the information it obtained when it was trained or obtain the policy from online sources (if the system supports it). In summary, the results revealed that BingAI consistently showed high proficiency across policies (with high reliability and consistency scores and low referencing scores). Bard displayed broader variabilities and pronounced inconsistencies (e.g., low referencing scores and marked variability in accuracy and completeness). ChatGPT's performance was a blend of high scores for certain combinations of questions and policies counterbalanced by stark inconsistencies across all criteria.



We observe that considering Uber's policy, ChatGPT-4's performance ranges between 10 and 100 scores in both question categories (Figure \ref{fig:iqr_adap}). Despite this variability, a strong median of 91.1 indicates its overall competence. The consistency was further emphasized by the narrow interquartile range (82 to 95.5). In the Spotify policy, all three categories saw the model reaching its zenith with maximum scores of 100. For Twitter and Airbnb, the medians (82 and 95.5, respectively for user-generated questions) were strong, and the compact interquartile ranges again indicated reliable performances. Facebook's policy showed a similar trend with a median above 70.75. ChatGPT-4 predominantly had scores close to 1 in Relevance across both question categories, with only a slight dip to 0.2 for $RC_{f2}$ (Figure \ref{fig:adaptive_combined}). Clarity remained fairly consistent, with many of its scores ranging between 0.8 to 1, but there was a notable drop to 0.1 for $RC_{f2}$. In terms of Accuracy, while GPT-4 generally performed well in user-generated questions, there was a clear reduction in its performance in FAQ questions, dropping as low as -0.7 and 0.3 in the $DM_{f2}$ and $RC_{f2}$ respectively. Completeness scores demonstrated a similar trend with higher scores in user-generated questions and diminishing results in the FAQ questions, the lowest being -0.2 for $RC_{f2}$. The Reference, however, remained relatively low throughout, with a peak score of 1 for $PD_e$ and a dip to -0.6 in several FAQ questions. Finally, ChatGPT demonstrated improved performance in handling policies with a higher proportion of questions on content not explicitly stated in the policy, particularly evident in the $SE_{f3}$ question for Twitter's policy.

\textbf{Bard}: Bard displayed wider variability than ChatGPT-4 (Figure \ref{fig:iqr_adap}). In the Uber policy, both the question categories witnessed a spread from 10 to 100, suggesting more variance in their responses. The broader interquartile range (43.7 to 9.1) compared to ChatGPT-4 underlined this. Twitter's performance indicated considerable inconsistency, with the lowest score being 10 and Q1 also at 10, suggesting that 25\% of responses were at the floor of the scoring metric. Airbnb's further echoed this inconsistency, with both minimum and Q1 at 3.7,4,6. However, Facebook policy in both question categories showed tighter interquartile ranges, hinting at a better consistency. Bard maintained a high Relevance, predominantly fluctuating between 0.4 to 1 in both category questions, but saw a drastic decline for  $DM_{f3}$, scoring 0.2 (Figure \ref{fig:adaptive_combined}). Its Clarity mostly mirrored ChatGPT-4's pattern, though it had a steeper drop in FAQ questions, reaching as low as 0.1 in the $DM_{f3}$ question. Accuracy exhibited significant variability, with scores ranging from a high of 0.7 in user-generated questions like $T_{u1}$ to a troubling -0.8 in $UC_{u1}$. For FAQ questions, the same variability was seen with 0.8 in $TU_{f3}$ as a high and -1 in $SE_{f3}$ as low. Completeness varied considerably as well, with scores peaking at 0.9 for $SE_{f1}$ and plummeting to -0.3 in $UC_{u1}$. Reference scores were particularly notable for Bard due to their consistent negative values, dropping as low as -1 for multiple questions, suggesting possible issues with citation or source integrity.

\textbf{BingAI}: BingAI showcased a peculiar trend (Figure \ref{fig:iqr_adap}). For Facebook's policy, it ranged from 37 to a perfect 100, with a commendable median of 95.5. Yet, the FAQ questions revealed stark contrasts from 10 to 100, with a median dropping to 88.7. This disparity between FAQ and user-oriented questions was further underscored by the interquartile range shift from 37-100 in FAQ questions to a much broader 46-86.5 in user-oriented questions. Similarly, Both question categories in Uber reflected a pronounced inconsistency, with both the minimum and 25\% of scores (Q1) languishing at 10, while the upper quartile (Q3) stretched to 82. Notably, for Airbnb, BingAI achieved an 84.25 median, indicating that over half of its responses received the maximum score, though its minimum at 10 demonstrates the presence of some extreme outliers. BingAI's performance in Relevance started strong, reaching 1 in categories like $T_{f2}$, $UC_{f1}$, and $PD_{f1}$, but faltered for $DM_{f2}$ question which scored -0.2 as shown in Figure \ref{fig:adaptive_combined}. Clarity remained relatively stable, with many scores hovering around the 0.6 to 1 range. However, its accuracy was inconsistent, dropping to -1 for $DM_{f3}$ but redeeming itself with scores like 1 in $PD_{f1}$. Completeness scores were highly variable, from a full score of 1 for $PD_{f1}$ to a concerning -1 for $DM_{f2}$. As for the Referencing, it scored negatively for most of the questions. Finally, BingAI exhibited its strongest performance in processing policies with a higher proportion of questions on not explicitly stated content.

\subsection{Assessing the Quality of Responses to Privacy Regulation Questions}

\begin{figure}[!htb]
\centering
\includegraphics[width=\columnwidth]{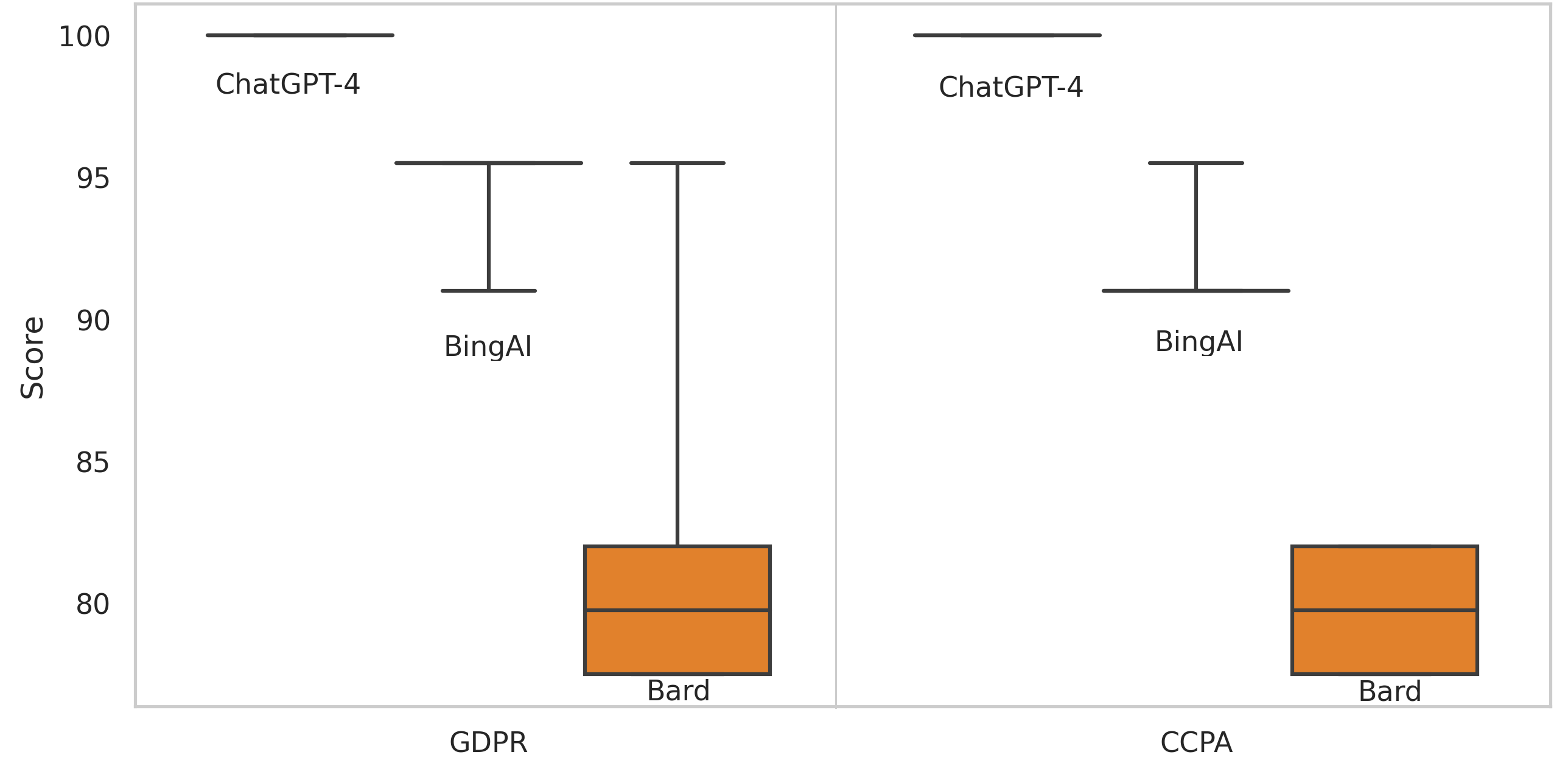}
\caption{Scores for privacy regulation questions.}
 
\label{fig:reg_box}
\end{figure}

This experiment examines the quality of responses generated by the systems for questions concerning the CCPA and GPDR data protection regulations. Figure \ref{fig:reg_box} shows the results obtained after executing the privacy regulation benchmark for both data protection regulations. Both ChatGPT-4 and BingAI excelled in answering privacy regulation queries, with ChatGPT-4 consistently achieving top scores across every metric. While Bard demonstrated good performance, it consistently struggled to provide accurate references, placing it behind the other two models.

For all six questions (i.e., $PR_1$, till $PR_{6}$), ChatGPT-4 and BingAI responses were accurate, relevant, comprehensive, and included correct references to regulation details. BingAI's scores took a hit due to its tendency to refer to online articles for its information rather than directly citing the articles from the GDPR and CCPA, a practice that ChatGPT-4 consistently followed. 
On the other hand, Bard's responses for questions $PR_1$, $PR_2$, and $PR_5$ scored 0.5 for completeness as they lacked some details. Also, the responses scored -1 across all questions wrt the reference metric for both CCPA and GDPR. 

\section{Discussion}
\label{sect:discussion}    
While, up to the author's knowledge, no specific GenAIPA has been proposed yet, our experiments indicate that current general-purpose genAI models can an effective tool when confronted with privacy-related questions. The three evaluated systems (i.e., Bard, BingAI, and ChatGPT-4) demonstrated commendable capabilities when evaluated with \benchmark. When addressing questions related to an organization's privacy policies, all systems obtained a fairly high score for all questions. In particular, BingAI emerged as the most consistent performer, demonstrating superior outcomes across most metrics. This is particularly interesting since BingAI and ChatGPT-4 use the same underlying model but are different chatbots which highlights the importance of developing robust layers that promote, for instance, appropriate referencing, on top of the genAI models. The three systems also showed a strong understanding of the two privacy regulations evaluated, scoring in general higher than for the privacy policies. This might be due to the fact that there has been more discussion about data privacy regulations online than about specific privacy policies. 

However, the three systems also encountered challenges in dealing with \benchmark. First, when paraphrased versions of the questions were used the performance of the systems reduced (particularly of BingAI). This inconsistency highlights that some systems might expect users to express their questions in specific ways, which would be an issue given the difference in perception about privacy among the general public~\cite{steijn2015privacy}. Second, of particular concern was a disconnect between the relevance and clarity of generated responses and their factual accuracy and completeness. Substantially incorrect responses were often presented coherently and relevantly, posing the risk of misleading users. Third, we observed frequent issues with respect to references that often point to outdated or incorrect data from the model's training set rather than the most recent privacy policy information (even when this policy was explicitly provided to the systems). 

We also explore which \benchmark questions seem to be easier and harder for the current genAI technology. To this end, we average the performance obtained for each question and their paraphrased versions for all the systems and policies. Table~\ref{tab:question_performance} summarizes the results with the top-5 questions with the highest and lowest score. In particular, we note that, in general, questions for which there is explicitly defined content in the policy tend to be easier for the systems. Notable exceptions were \( RC_{u1} \), which obtained a high score in spite of all the policies lacking information about whether any privacy breach had occur, and \( CA_{f1} \) which obtained a lower score in spite of all policies mentioning their compliance to GDPR/CCPA and/or other regulations. This might be explained because of the tendency of genAI systems to create non existing content. It is also worth noting that questions about transparency and those explicitly posed by individual users seem to be easier for the technology, especially when compared to questions on compliance and accountability topics that obtained lower scores in general.

\begin{table}[h]
\centering
\small
\begin{subtable}{.5\linewidth}
\centering
\begin{tabular}{|l|l|p{1.1cm}|}
\hline
\textbf{Q} & \textbf{Score} & \textbf{Missing Content} \\ \hline
\( T_{u1} \)     & 8.99           &  None                     \\ \hline
\( DM_{u1} \)    & 8.58           &  None                     \\ \hline
\( T_{f2} \)     & 8.57           & None                     \\ \hline
\( CA_{u1} \)    & 8.40           & None                     \\ \hline
\( RC_{u1} \)    & 8.4            & All                      \\ \hline
\end{tabular}
\caption{ }
\label{subtab:top_question_performance}
\end{subtable}%
\begin{subtable}{.5\linewidth}
\centering
\begin{tabular}{|l|l|p{1.1cm}|}
\hline
\textbf{Q} & \textbf{Score}  & \textbf{Missing Content} \\ \hline
\( DM_{f3} \)     & 4.23           & A, U, S    \\ \hline
\( T_{f3} \)      & 5.2            & F, S, A\\ \hline
\( AEE_{f2} \)    & 5.91           &  T, F        \\ \hline
\( CA_{f3} \)     & 5.95           &  T, F, S, A \\ \hline
\( CA_{f1} \)     & 6.33           & None                     \\ \hline
\end{tabular}
\caption{ }
\label{subtab:bottom_question_performance}
\end{subtable}
\caption{Top-5 highest (a) and lowest (b) scoring questions across systems.}
\label{tab:question_performance}
\end{table}

We also analyze which privacy policies seem to be easier or harder to process by current genAI technology. To this end, we average the score obtained for all metrics, questions (including their paraphrased versions), and systems for each privacy policy. We observe that the highest scores are obtained for the privacy policies of Facebook and Airbnb and the lowest for the privacy policies of Uber and Spotify. Note that, while the percentage of unique and connective words are similar across policies, the main differences between the policies are observed with respect to their length and required reading level. The ranking obtained points to the length of the policy being a decisive factor in performance. In particular, the highest and lowest score is obtained by the shortest and longest policies (Facebook's and Uber's, respectively). Interestingly, even when the reading level of the Airbnb policy was two levels above the Uber policy, it scored higher which indicates that while it is a critical factor for users to understand the policy, it might not be for the genAI systems.

\section{Conclusion and Future Work}
\label{sect:future}
The emergence of generative AI systems and their ability to summarize text and answer questions generating human-like text presents an opportunity to develop more sophisticated privacy assistants (GenAIPAs). Due to the implications for individuals receiving wrong information that might impact their privacy, it is required to evaluate such systems properly. In this paper, we have presented a benchmark, \benchmark, to evaluate future GenAIPAs, which includes questions about privacy policies and data privacy regulations, evaluation metrics, and annotated privacy documents. Our evaluation of popular genAI technology, including ChatGPT, Bard, and BingAI, shows promise for the technology but highlights that significant work remains to enhance their capabilities in handling complex queries, ensuring accuracy, maintaining response consistency and citing proper sources. One limitation of this paper, is that we included only policies and questions in English. As future work, we plan to continue expanding \benchmark with more annotated answers for a larger number of privacy documents (and in multiple languages) to maintain its relevance and utility. We also aim to develop the infrastructure to perform a periodic evaluation of current and future versions of genAI and GenAIPA systems.

\bibliographystyle{ieeetr}
\bibliography{References} 

\begin{appendices}

\section{Evaluator}
\label{sect:evaluator}

\benchmark includes a component evaluator whose goal is to communicate with the GenAIPA sharing the privacy documents and questions and collecting answers and summaries (see Algorithm \ref{alg:revised_privacy_doc_analysis}). The methodology centers around evaluating responses to a set of questions based on a privacy document (PD) and a company name (CN). The procedure unfolds over multiple iterations, each comprising different initializations that focus on either the PD or the CN. This approach is tailored to examine how the GenAIPA system's responses vary under distinct contextual setups. For each run $i$ in the total number of runs $r$, the procedure undertakes the following distinct initialization:
\begin{itemize}
\item\textbf{Initialization with Company Name (CN):} The GenAIPA is introduced to the CN, forming the context for the subsequent query execution. This approach is designed to assess how the system interprets and responds to questions when primed with the company name. Here, the evaluator engages the GenAIPA by explaining that it will pose questions about the privacy policy of a specific organization, such as Uber. The purpose of this approach is to determine whether GenAIPA possesses prior knowledge about the company's privacy policy and to gauge the accuracy of its responses based solely on this knowledge. This evaluation aspect is crucial for understanding the extent to which the GenAIPA can rely on its pre-existing data in generating informed responses about a company's privacy practices.
\item\textbf{Initialization with Privacy Document (PD):} In this phase, the GenAIPA attention is directed towards the PD. The system is introduced to the specific content within the privacy document. This method is pivotal for analyzing how effectively the GenAIPA can generate responses that are directly influenced by the detailed information provided in the PD. To accommodate potential token limit constraints of GenAIPAs, the initial prompt clarifies that the privacy document will be delivered in segmented portions. This approach ensures that the GenAIPA comprehensively processes the document in manageable segments. Following the introduction of each segment, the system is then presented with questions related to the content of the PD. This step is critical for assessing the GenAIPA's capability to understand and respond accurately to queries that are directly tied to the nuances and specificities of the privacy document. 
\item\textbf{Initialization with Summary based on CN:} The procedure also involves a unique initialization where GenAIPA is requested to summarize the privacy document based on CN, albeit without an explicit introduction to the PD. This step is designed to gauge the AI system's capacity to synthesize and summarize content based on its pre-existing knowledge or understanding. The initial prompt directs GenAIPA to create a summary, thereby providing a foundation for subsequent queries. This method tests the system's ability to process and condense information in the absence of direct exposure to the full document, focusing on its internal processing capabilities and prior knowledge.
\item\textbf{Initialization with Summary based on PD:} Similar to the previous step, but this time the summary is generated based on the PD. This initialization tests the system's response efficiency when working with a condensed version of the privacy document.
\end{itemize}

\paragraph{Query Execution and Data Collection: }
In each initialization, the set of questions $Q$ is shuffled (as $Q'$) to introduce variability. The system executes these queries, and the responses are collected. For each initialization method, the responses are stored separately, identified as $A1,A2,A3, and A4$.
Aggregation of Responses
Upon completion of all initializations for a single run, the responses from each method ( $A1,A2,A3, and A4$) are aggregated into a comprehensive list $A$. This process is repeated for each run, enriching $A$ with a diverse set of responses that reflect the system's performance across different contexts.
The algorithm concludes by returning the aggregated data $A$, which encompasses the varied responses generated under each initialization scenario. This output serves as a valuable dataset for analyzing the GenAIPA's adaptability and accuracy in responding to privacy-related inquiries under different contextual influences.

\begin{algorithm}[ht]
\caption{Revised Privacy Document Analysis and Query Response} 
\label{alg:revised_privacy_doc_analysis}
\begin{algorithmic}[1]
\Procedure{GenerateAndStoreQueryResponses}{Privacy Document PD, Company Name CN, Questions Q, Runs r}
\State Initialize $A$ as an empty list
\For{$i=1$ to $r$}
    \State $Q' \gets \text{ShuffleQuestions}(Q)$
    \State \textit{\#initialization 1}
    \State IntroducePrivacyDocument(CN)
    \State $A1 \gets \text{QueryExecution}(Q')$
    \State ResetConversation()

    \State \textit{\#initialization 2}
    \State IntroducePrivacyDocument(PD)
    \State $A2 \gets \text{QueryExecution}(Q')$
    \State ResetConversation()

    \State \textit{\#nitialization 3}
    \State IntroducePrivacyDocument(CN)
    \State $S \gets \text{GenerateSummary}$
    \State ResetConversation()
    \State IntroducePrivacyDocument(S)
    \State $A3 \gets \text{QueryExecution}(Q')$
    \State ResetConversation()

    \State \textit{\#initialization 4}
    \State IntroducePrivacyDocument(PD)
    \State $S \gets \text{GenerateSummary}$
    \State ResetConversation()
    \State IntroducePrivacyDocument(S)
    \State $A4 \gets \text{QueryExecution}(Q')$
    \State ResetConversation()

    \State $A \gets A + [A1, A2, A3, A4]$
\EndFor
\State \textbf{Return} $A$
\EndProcedure
\end{algorithmic}
\end{algorithm}

\begin{algorithm}[H]
\caption{GenAIPA Response Evaluation}\label{alg:policy_evaluation3}
\begin{algorithmic}[1]
\Procedure{EvaluateResponse}{Scores A provided by analyst}
\State $P \gets \emptyset$
\For{$i = 1$ to $|A[1]|$} 
\State $\overline{A_i} \gets \frac{1}{|A|} \sum_{score \in A} score[i]$
\State Categorize $\overline{A_i}$ as Green, Yellow, or Red
\State $P \gets P \cup {\text{Category}}$
\EndFor
\State \textbf{return} $P$
\EndProcedure
\end{algorithmic}
\end{algorithm}
In the GenAIPA Response Evaluation process (see Algorithm \ref{alg:policy_evaluation3}), the analyst scrutinizes the generated responses based on five key features: Relevance, Accuracy, Clarity, Completeness, and Reference to policy sections. Scores for each feature are provided as input $A$. The procedure starts by initializing an empty set $P$ and then iterates through each score $A_i$ in $A$, accumulating them for subsequent categorization. This evaluation is performed for each set of scores, representing multiple runs, and the average scores for each run are determined. Ultimately, an overall average score is calculated across all runs. This average score is then categorized into one of three groups: Green, Yellow, or Red, based on predefined criteria (see Section \ref{sect:metrics}. The category is stored in set P, which upon completion of all iterations, contains the categorized average scores for all runs. The primary goal of this evaluation is to offer insights into GenAIPA's capabilities in generating privacy policy-related responses and identify areas of potential improvement.

\section{Additional Experiments}

This section includes additional experiments and results from our evaluation of ChatGPT-4, Bard, and BingAI using \benchmark.

\subsection{Assessing the Quality of Privacy Policy Summaries} 

\begin{figure*}[!htb]
     \centering
     \begin{subfigure}[b]{0.99\textwidth}
         \centering
         \includegraphics[width=\textwidth]{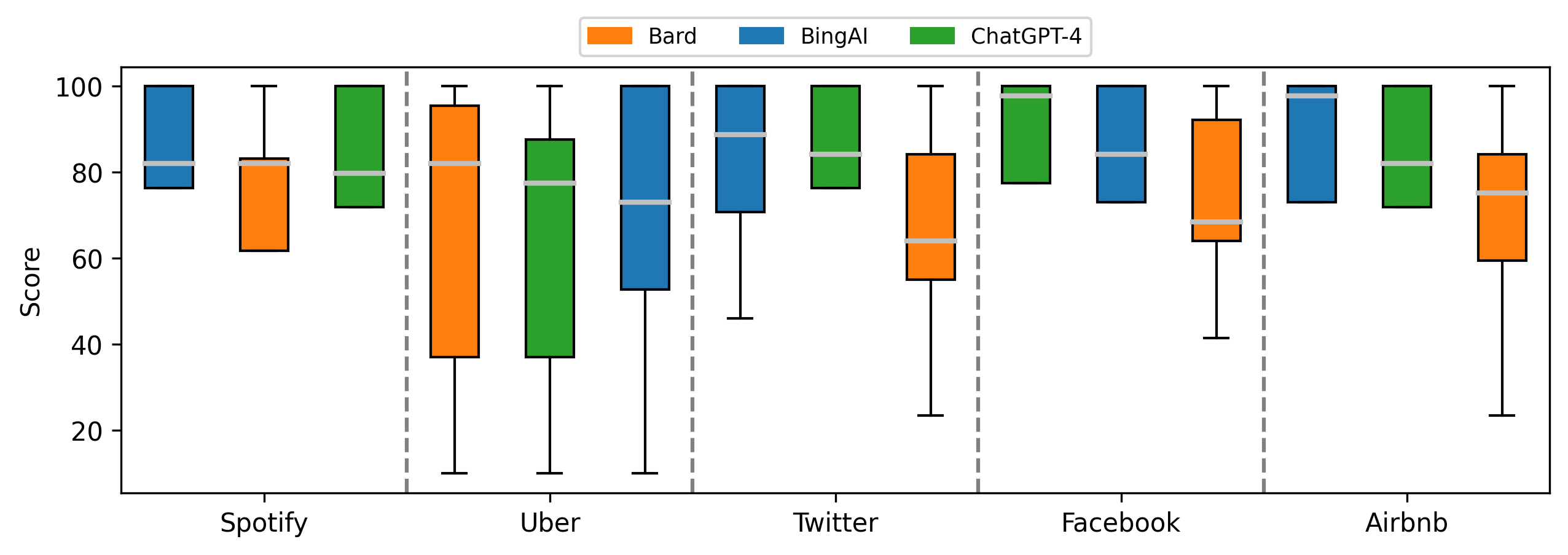}
          \caption{Overall score per policy.}
         \label{fig:iqr_summary}
     \end{subfigure}
     \hfill
     \begin{subfigure}[b]{0.99\textwidth}
         \centering
         \includegraphics[width=\textwidth]{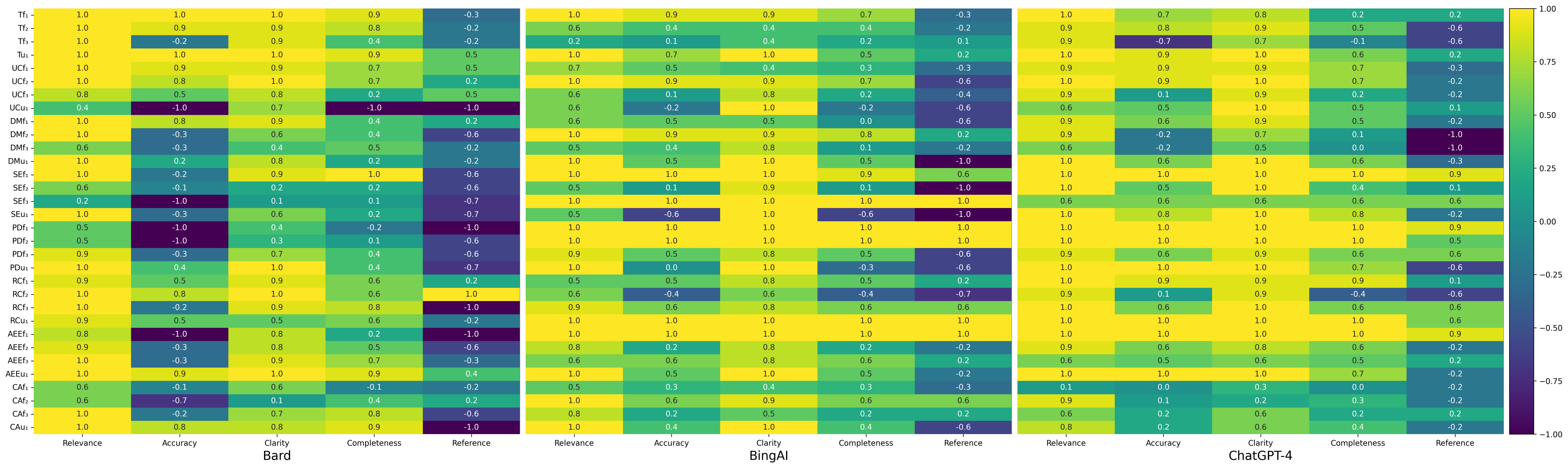}
          \caption{Average scores for all policies across metrics.}
         \label{fig:summary_combined}
         \caption{Performance of systems when the privacy policy summary is explicitly shared.}
     \end{subfigure}
     
     
\end{figure*}

This experiment aims to examine the quality of the summary generated for privacy policies. 
    The results (see Figure \ref{fig:sum_over}) demonstrate that ChatGPT-4 consistently excelled across various tasks, particularly in handling policies with higher non-existing content. Bard, showing promise in simpler FAQ questions, faced challenges in more complex, user-generated content. BingAI, adaptable yet exhibiting performance variability, performed best in medium-difficulty challenges but sometimes struggled with complex questions. Both Bard and BingAI encountered difficulties in citing references accurately, underscoring areas for improvement in understanding, clarity, consistency, and accuracy.

\textbf{ChatGPT-4}:
ChatGPT-4 demonstrated remarkable consistency across diverse policies (see Figure \ref{fig:iqr_summary}). In Spotify's policy, with a moderate level of non-existing content (25\%), the model achieved a median score of 79.75. This performance slightly dipped in the more complex Uber policy, which has a similar percentage of non-existing content but a longer reading time, resulting in a median of 77.5. In Twitter's policy, characterized by a significant amount of non-existing content (43.7\%), ChatGPT-4 maintained a strong median of 84.25. Its performance peaked in Facebook's policy, with the highest non-existing content (46.8\%), achieving an impressive median of 97.75. For Airbnb, despite the policy's high reading level (14.15 FKGL), ChatGPT-4 upheld a solid median of 82, showcasing its adaptability to complex information. Figure \ref{fig:summary_combined} reveals ChatGPT-4's robust performance across different metrics. It achieved high scores in Relevance, particularly in FAQ questions such as '\(SE_{f1}\)' and '\(PD_{f2}\)' (both scoring 1.0), signifying its effective understanding and summarization capabilities. However, challenges were noted in Clarity and Completeness for user-generated questions like '\(DM_{f3}\)' (scoring -1.0) and '\(UC_{f2}\)' (scoring -0.2), indicating some difficulty in maintaining coherence and thoroughness. Accuracy fluctuated, performing strongly in questions like '\(SE_{f1}\)', '\(PD_{f1}\)', and '\(PD_{f2}\)' (all scoring 1.0), but showing weaknesses in '\(T_{f3}\)' (-0.7) and '\(DM_{f2}\)' (-0.2). Completeness varied, ranging from high scores in '\(SE_{f1}\)' and '\(PD_{f1}\)' (1.0) to lower scores in '\(RC_{f2}\)' (0.4). Reference metrics, though generally strong in areas like '\(SE_{f1}\)' and '\(PD_{f1}\)' (0.9), revealed potential areas for improvement, particularly in '\(DM_{f2}\)' and '\(DM_{f3}\)' (both scoring -1.0), suggesting a need to enhance source citation accuracy.

\textbf{Bard}: Bard's performance, as indicated by the IQR data (see Figure \ref{fig:iqr_summary}), varied significantly across different policies. For Spotify, it achieved a median of 82, suggesting competent handling of standard policy content. However, in the more complex Uber policy, the model displayed a wider performance range (Min: 10, Q3: 95.5), indicating inconsistencies in handling diverse and challenging content. Twitter's policy, with a large amount of non-existing content, was more challenging for Bard, resulting in a lower median score of 64. In Facebook's policy, Bard managed a median score of 68.5, showing some capability in dealing with incomplete information but also highlighting room for improvement. Airbnb's policy posed a moderate challenge, with Bard achieving a median score of 75.25. Figure \ref{fig:summary_combined} presents a mixed performance across various metrics. In Relevance, Bard scored high in simpler FAQ questions such as '\(T_{f1}\)' and '\(Tu_{f1}\)' (both scoring 1.0), but it struggled with more nuanced user-generated questions, evidenced by lower scores in '\(SE_{f3}\)' (0.2). Clarity was similarly variable, with high scores in '\(T_{f1}\)' and '\(Tu_{f1}\)' (1.0), but significantly lower scores in more complex questions like '\(PD_{f2}\)' (0.3) and '\(CA_{f2}\)' (0.1). This inconsistency in Clarity, especially in more complex or nuanced questions, underscores a need for Bard to improve its ability to convey information clearly and effectively. Accuracy showed similar fluctuations, with top scores in straightforward questions like '\(T_{f1}\)' and '\(Tu_{f1}\)' (1.0), but it faltered in questions like '\(UCu_{f1}\)', '\(PD_{f1}\)', and '\(PD_{f2}\)' (all scoring -1.0). This variability in Accuracy, particularly in user-generated questions, suggests Bard's potential challenges in consistently maintaining precision. Completeness spanned from high scores in '\(T_{f1}\)' and '\(SE_{f1}\)' (1.0) to lows in '\(UCu_{f1}\)' and '\(PD_{f1}\)' (-1.0), indicating fluctuating thoroughness in its responses. Reference metrics were notably poor, with about half of the questions scoring below -0.6, indicating a significant area for improvement in citing sources and maintaining informational integrity.

\textbf{BingAI}: In the IQR data analysis, BingAI showed distinctive performance patterns across policies. Facebook's policy, with high non-existing content (46.8\%), saw BingAI range from 28 to 100, achieving a median score of 84.25, demonstrating its adaptability to varying content within the same policy. However, Uber's policy presented a challenge, particularly in user-generated questions, where BingAI's median score was only 73, reflecting the difficulty in achieving uniformity and consistency. In Airbnb's policy, characterized by substantial non-existing content and a high FKGL level, BingAI excelled with a median score of 97.75. It also performed effectively in Twitter's policy and Spotify's policy, attaining median scores of 88.75 and 82, respectively. Figure \ref{fig:summary_combined} showcases its performance across different metrics. In Relevance, BingAI scored highly in FAQ questions like '\(SE_{f1}\)', '\(PD_{f1}\)', and '\(PD_{f2}\)' (all scoring 1.0), reflecting its strong understanding and summarization of policy content. However, it faced challenges in '\(T_{f3}\)' (scoring 0.2), indicating areas where it could enhance its comprehension. Clarity was generally good, with high scores in '\(SE_{f1}\)' and '\(PD_{f1}\)' (1.0), but slightly lower in '\(DM_{f1}\)', '\(T_{f2}\)', '\(T_{f3}\)', and '\(CA_{f1}\)' (0.4), suggesting some variability in presenting information clearly. Accuracy showed strengths and weaknesses, with high scores in '\(SE_{f1}\)' and '\(PD_{f1}\)' (1.0) but lower scores in '\(SEu_{f1}\)' (-0.6) and '\(RC_{f2}\)' (-0.4), indicating areas where BingAI could improve in maintaining factual correctness. Completeness varied, scoring high in '\(SE_{f3}\)' and '\(PD_{f1}\)' (1.0) but showing lower scores in '\(SEu_{f1}\)' (-0.6) and '\(RC_{f2}\)' (-0.4), highlighting inconsistency in covering all relevant aspects of the policy content. Reference metrics were notably weak, with 75\% of the questions scoring below 0, pointing to a significant area for improvement in providing well-cited and reliable summarizations.

In conclusion, while GenAIPAs demonstrate promising capabilities in summarizing privacy policies and answering specific privacy-related questions, their effectiveness is closely tied to the availability and complexity of content within these policies. ChatGPT-4 distinguishes itself with consistent performance and adaptability, particularly in handling policies with higher non-existing content. Bard shows promise in simpler FAQ questions but faces challenges with more complex user-generated content, indicating a need for enhanced understanding and clarity. BingAI, while adaptable, exhibits variability in its performance, especially in user-generated questions, suggesting areas for further improvement in consistency and accuracy.

\end{appendices}
\end{document}